\documentclass{acmart}
\newcommand{\modelnamenospace}{PCR}
\newcommand{\modelname}{PCR }
\AtBeginDocument{%
  }
\usepackage{subfig}
\usepackage{multirow}
\usepackage{caption}
\usepackage{graphicx}
\usepackage{algorithm}        
\usepackage{algpseudocode}   
\setcopyright{acmlicensed}
\copyrightyear{2018}
\acmYear{2018}
\acmDOI{XXXXXXX.XXXXXXX}
\acmConference[Conference acronym 'XX]{Make sure to enter the correct
  conference title from your rights confirmation email}{June 03--05,
  2018}{Woodstock, NY}
\acmISBN{978-1-4503-XXXX-X/2018/06}
\begin{document}

\title{PCR: A Prefetch-Enhanced Cache Reuse System for Low-Latency RAG Serving}




\author{Wenfeng Wang}
\affiliation{
  \institution{Shanghai Jiao Tong University}
  \city{Shanghai}
  \country{China}
}

\author{Xiaofeng Hou}
\authornote{Corresponding author.}
\affiliation{
  \institution{Shanghai Jiao Tong University}
  \city{Shanghai}
  \country{China}
}

\author{Peng Tang}
\affiliation{
  \institution{Shanghai Jiao Tong University}
  \city{Shanghai}
  \country{China}
}

\author{Henyi Zhou}
\affiliation{
  \institution{Shanghai Jiao Tong University}
  \city{Shanghai}
  \country{China}
}

\author{Jing Wang}
\affiliation{
  \institution{East China Normal University}
  \city{Shanghai}
  \country{China}
}

\author{Xinkai Wang}
\affiliation{
  \institution{Shanghai Jiao Tong University}
  \city{Shanghai}
  \country{China}
}

\author{Chao Li*}
\affiliation{
  \institution{Shanghai Jiao Tong University}
  \city{Shanghai}
  \country{China}
}

\author{Minyi Guo}
\affiliation{
  \institution{Shanghai Jiao Tong University}
  \city{Shanghai}
  \country{China}
}

\ccsdesc[500]{Computer systems organization~Memory architectures}
\ccsdesc[500]{Computer systems organization~Embedded systems}
\ccsdesc[300]{Computing methodologies~Machine learning approaches}
\ccsdesc[300]{Computing methodologies~Neural networks}
\ccsdesc[100]{Computer systems organization~Real-time systems}

\begin{abstract}

Retrieval-Augmented Generation (RAG) systems enhance the performance of large language models (LLMs) by incorporating supplementary retrieved documents, enabling more accurate and context-aware responses. However, integrating these external documents often results in very long input sequences, which significantly increases computation costs during the prefill stage, where key-value (KV) representations for all input tokens are generated. This latency bottleneck becomes especially pronounced under high-throughput serving scenarios.
KV-cache reuse offers a promising solution by storing previously computed KV states for shared input prefixes, thereby avoiding redundant computation across requests that contain overlapping context. Yet, the effectiveness of cache reuse is often limited by three practical challenges: low cache hit rates due to naive eviction policies, high CPU–GPU data transfer overhead, and slow SSD I/O when caches spill to storage.
To address these issues, we propose \modelnamenospace, a system designed to maximize KV-cache reuse efficiency through intelligent prefetching and pipelined data movement. Specifically, \modelname introduces three key techniques: (1) a prefix-tree caching structure with a look-ahead LRU replacement policy that uses pending requests in the scheduler queue to improve cache hit ratios; (2) layer-wise overlapping that pipelines KV-cache loading and GPU computation across CUDA streams to hide communication latency; and (3) queue-based prefetching that proactively loads relevant KV caches from SSD into DRAM before they are needed. Extensive experiments show that \modelname outperforms existing KV-cache reuse methods, achieving up to a 2.47× speedup in terms of average TTFT.

\end{abstract}

\maketitle
\begingroup
\renewcommand\thefootnote{}
\footnotetext{New Paper, Not an Extension of a Conference Paper.}
\endgroup

\section{Introduction}

Recently, numerous large language models (LLMs) and tailored optimizations have emerged, demonstrating impressive performance across various domains~\cite{guo2025deepseek, achiam2023gpt, grattafiori2024llama, tang2024hobbit, liu2024survey, wang2025moe}. To further enhance their effectiveness in domain-specific tasks, Retrieval-Augmented Generation (RAG) has been introduced, which provides LLMs with supplementary domain-specific information, thereby improving their task comprehension~\cite{gao2023retrieval, zhao2024retrieval, lewis2020retrieval, guu2020retrieval}. With RAG, LLMs can efficiently tackle new tasks without requiring additional training, significantly reducing training costs. Consequently, RAG has found applications in diverse scenarios, such as chatbots and document summarization, enhancing user experience.

Traditionally, RAG comprises two primary stages, including retrieval and generation. When a request arrives, the retriever first queries a database to locate the most relevant documents, which are then combined with the original request to create an enhanced input for the generator (e.g., LLM). Leveraging this enriched context, the generator can produce responses that are more informative and accurate, significantly improving the quality compared to responses that are just based on the original request alone. 


Combining additional documents with the original input can improve the quality of responses, but also  resulting in very long sequences for LLMs to process. This significantly increases latency in the prefill stage (Time to First Token, TTFT), often dominates the whole inference process when input length is very long and the decoding length is relatively short~\cite{ragcache}. What's more, as shown in Figure~\ref{fig:kv-stroage-ttft}, TTFT  grows super-linearly with the input length due to the computing bound of GPU in prefill stage.  Therefore, optimizing the prefill stage in RAG system is crucial to accelerate the overall inference speed and provide great service to users.

\begin{figure}[t]
    \centering
    \includegraphics[width=0.75\linewidth]{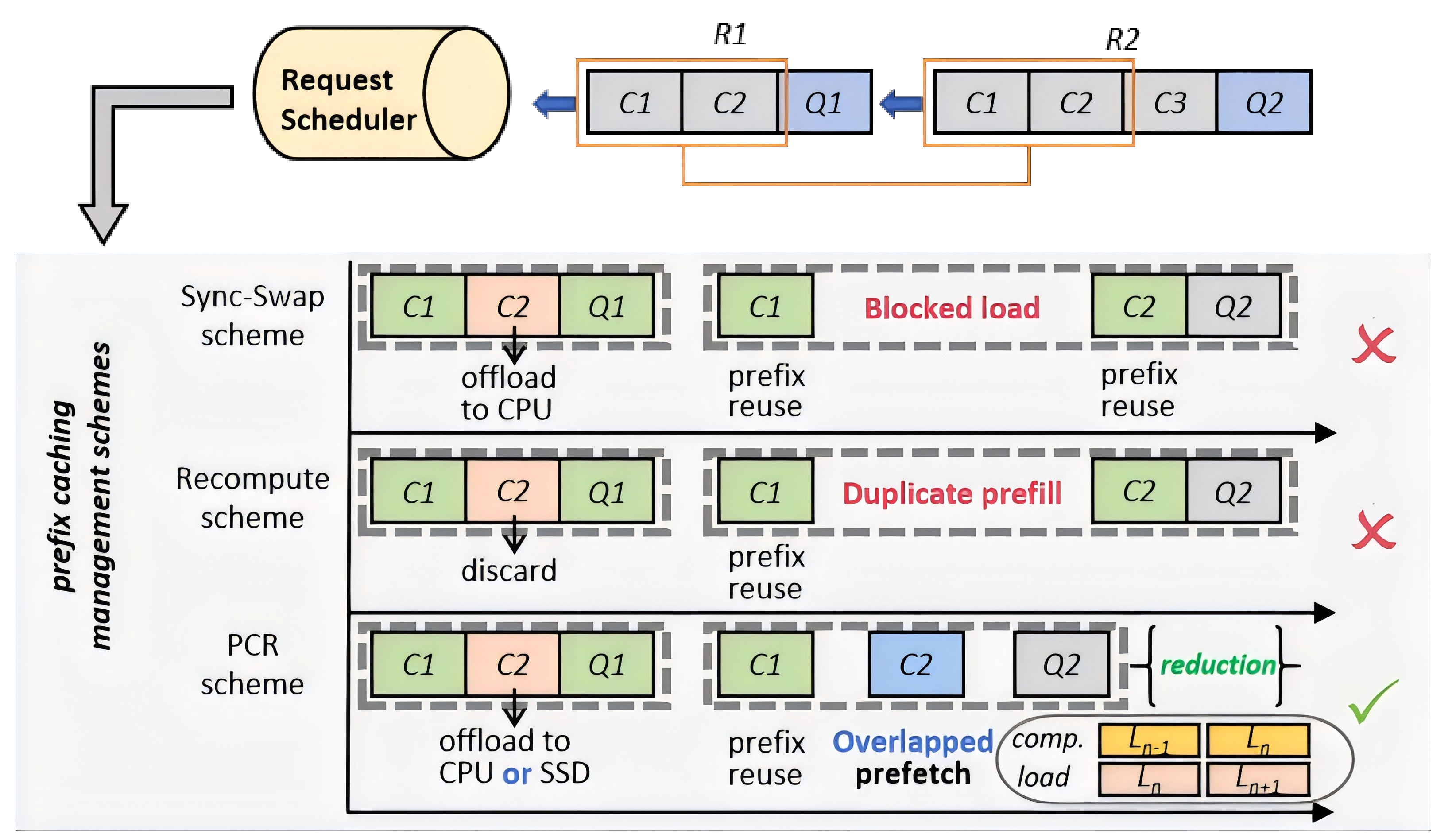}
    \caption{Comparison of prefix caching management schemes in LLM serving.}
    \label{fig:motivation}
    \Description{Motivation plot.}
\end{figure}


To reduce TTFT, recent research primarily focuses on KV-cache reuse, exploiting the frequent recurrence of identical documents across different LLM inputs. Approaches such as RAGCache~\cite{ragcache} and PromptCache~\cite{promptcache} utilize CPU memory to store KV caches generated from prior sequences, enabling reuse by reloading them into GPU memory when matches are found. However, the limited capacity of CPU memory constrains the scalability of such methods, especially when dealing with large RAG databases. Moreover, the loading and offloading overhead between CPU and GPU often offsets the potential latency reduction benefits of KV-cache reuse. Other works, including CacheGen~\cite{liu2024cachegen} and CacheBlend~\cite{cacheblend}, aim to reduce KV-cache reuse overhead by compressing or selectively skipping KV caches. While these approaches can improve efficiency, they often compromise model accuracy and are less effective for complex or diverse input scenarios.

As illustrated in Figure~\ref{fig:motivation}, existing prefix caching management schemes struggle to maintain efficiency when GPU memory capacity is exhausted. The Recompute scheme aggressively discards inactive cache blocks (e.g., C2) when system memory is exhausted, which inevitably leads to expensive duplicate prefill computations when those shared prefixes are encountered in subsequent requests. Alternatively, the Sync-Swap scheme mitigates recomputation by offloading evicted caches to host memory; however, it introduces severe blocked loads during synchronous retrieval, leaving the GPU idle and stalling the computational pipeline. To overcome these inherent bottlenecks, our work focuses on prefix-based cache reuse to maintain model accuracy, while simultaneously designing a prefetch-enhanced system to mitigate KV-cache reuse overhead through asynchronous loading from CPU memory and SSD.

First, to enable efficient prefix KV-cache reuse, we divide long inputs into fixed-size chunks and organize them using a prefix tree for fast matching, where each child node depends on its parent because the KV cache is position-dependent. To further improve the cache hit ratio, we propose a look-ahead LRU cache replacement policy, which leverages the information from upcoming requests in the waiting queue to prioritize chunks that will be reused soon.

Second, when reusing KV caches stored in CPU memory, the system must load matched KV caches and offload newly generated ones, which can stall the LLM and increase latency. Given the layer-wise structure of both LLMs and their KV caches, we introduce layer-wise overlapping, which hides the loading and offloading overhead by executing computation, loading, and offloading across three independent streams. This overlap reduces the effective overhead to only the first-layer loading and the last-layer offloading.

Third, when extending cache storage to SSDs, the slower read speed can introduce additional latency that cannot be fully hidden by computation. To address this, we design a queue-based prefetching mechanism that preloads the KV cache from SSD to CPU memory using the information from the waiting queue. Specifically, while the current request is being processed by the LLM, a dedicated thread simultaneously prefetches the KV caches for future requests. As a result, when these requests are processed, their KV caches can be directly loaded from CPU memory instead of SSD, significantly reducing latency.

In summary, our main contributions are as follows:

\begin{itemize}
\item A prefix-tree caching scheme equipped with a look-ahead LRU replacement policy to manage the KV-cache storage and improve KV-cache hit ratio.
\item A layer-wise overlapping technique to reduce CPU–GPU communication overhead by overlapping GPU computation with KV-cache transferring.
\item A queue-based prefetching mechanism to reduce SSD loading latency via asynchronous KV-cache prefetching.
\item An implementation on top of vLLM~\cite{kwon2023efficient}, evaluated on Llama and Qwen models, demonstrating an average 15\% reduction in TTFT compared with state-of-the-art systems.
\end{itemize}



\section{Background}
\begin{figure}[t]
    \centering
    \includegraphics[width=0.75\linewidth]{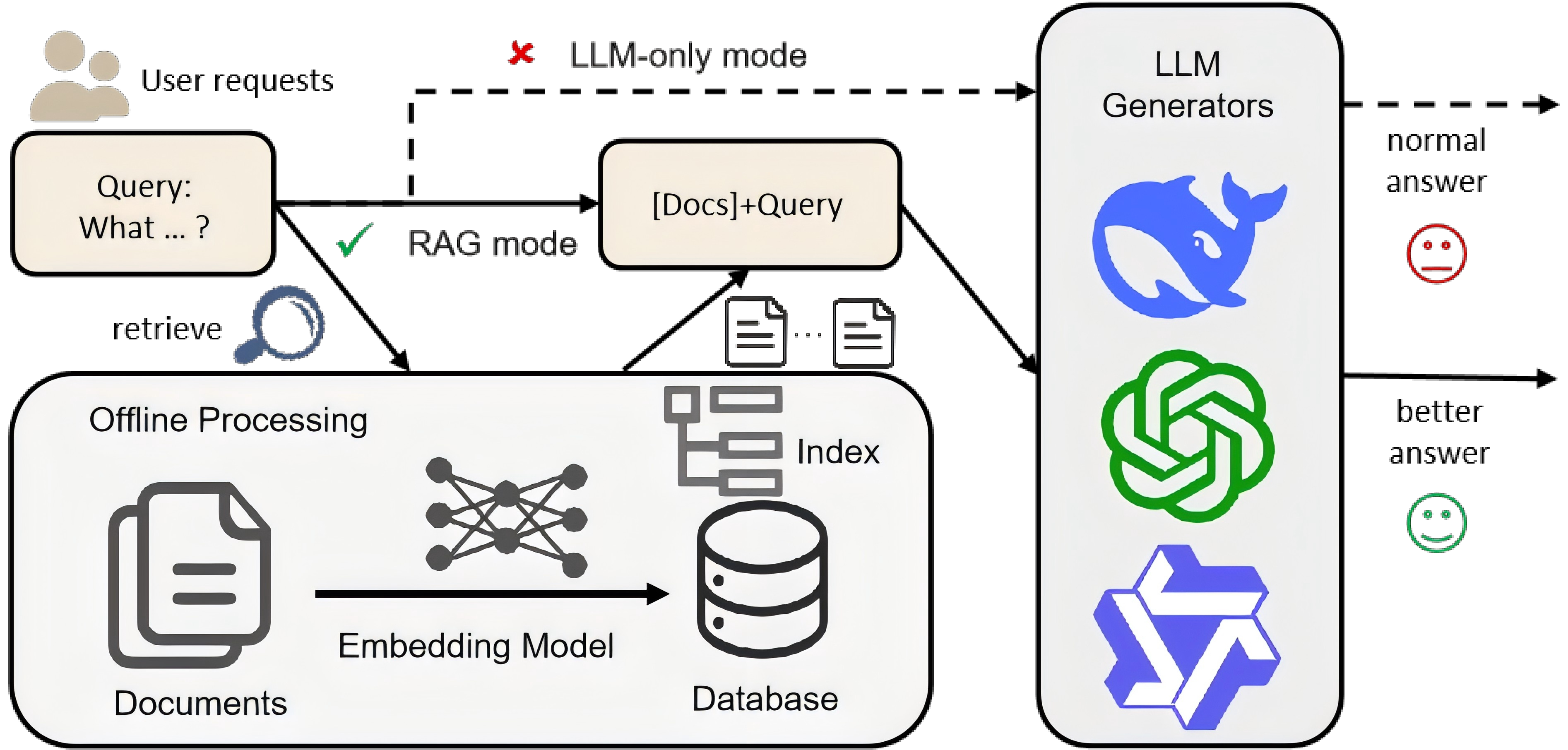}
    \caption{The workflow of RAG system.}
    \label{fig:rag-workflow}
    \Description{workflow plot.}
\end{figure}

\subsection{Retrieval-augmented Generation}
 
In the traditional LLM-only mode, the language model generates answers based solely on the knowledge it acquired during training. However, since information constantly evolves, the LLM may not reflect the most up-to-date knowledge. Moreover, it typically lacks access to domain-specific or private information that wasn't included in its training data. While retraining or fine-tuning the LLM can update its knowledge, this process is resource-intensive and costly. RAG addresses these limitations by combining the language understanding capabilities of LLMs with real-time knowledge retrieval~\cite{izacard2021leveraging}. As illustrated in Figure~\ref{fig:rag-workflow}, a RAG system consists of two main stages: Offline Processing and Online Inference.


In the offline stage, the goal is to construct a database containing materials for knowledge augmentation. This process includes segmenting documents into chunks, assigning unique indices, and converting them into vector representations using an embedding model. These vectors are then indexed using efficient similarity search algorithms, such as Approximate Nearest Neighbor  methods~\cite{indyk1998approximate}. In practice, many mature libraries, such as Faiss~\cite{douze2024faiss} and HNSW~\cite{malkov2018efficient}, can be employed to efficiently build and manage the retrieval database.

In the online stage, when a query is received, it is first embedded into a vector using the same model employed during database construction. The query vector is then compared against the indexed database to retrieve the most relevant document vectors, which are subsequently mapped back to their corresponding document chunks. The retrieved documents are concatenated with the query and provided as input to the LLM. By leveraging this supplemental information, the LLM can generate responses that are more accurate and contextually grounded compared to using the query alone.


Because the database is maintained independently of the LLM, it can be updated with new documents without the need to retrain the model, significantly reducing maintenance costs. Additionally, RAG helps reduce hallucinations by grounding the LLM's responses in actual source material.


\begin{figure}[t]
    \centering
    \includegraphics[width=0.9\linewidth]{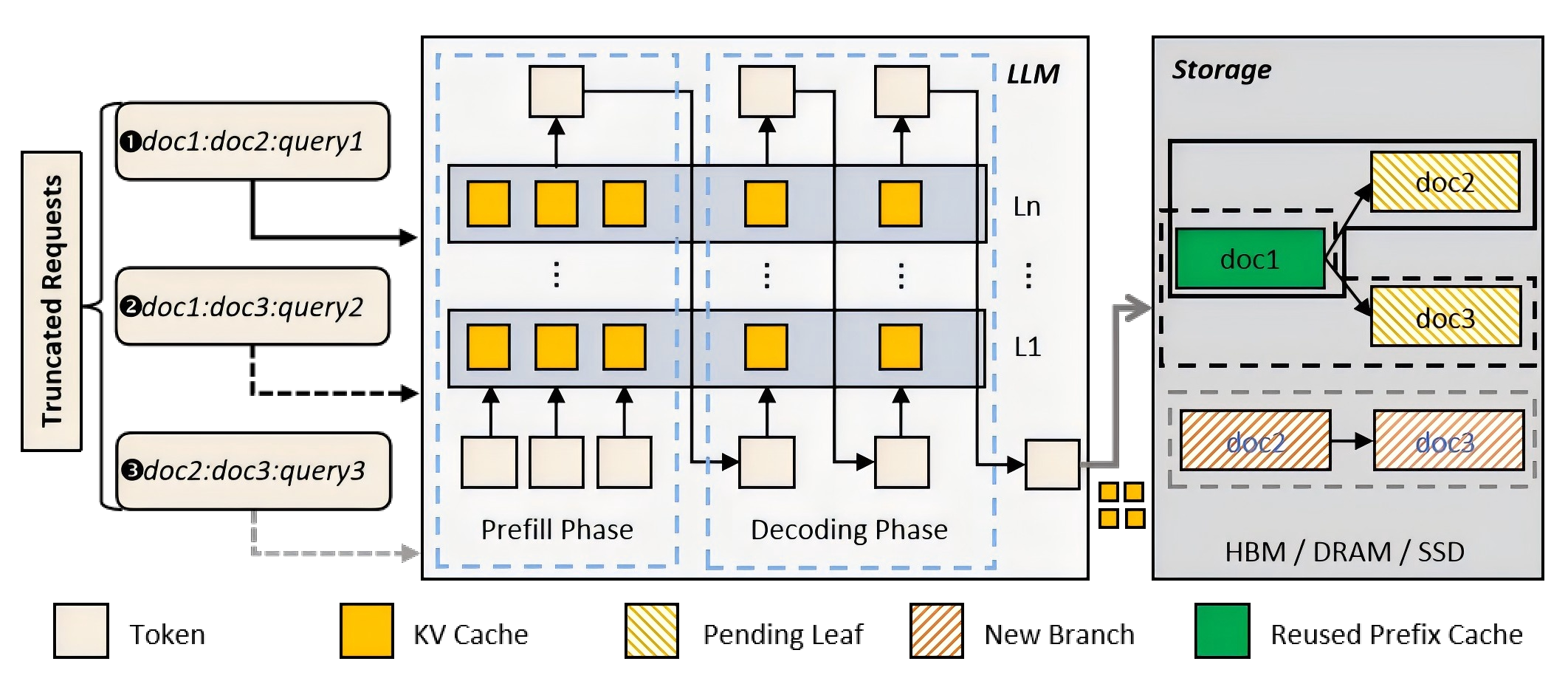}
    \caption{The process of prefix KV-cache reuse.}
    \label{fig:prefix-workflow}
    \Description{Prefix reuse plot.}
\end{figure}

\subsection{Prefix KV Cache Reuse}

As illustrated in Figure~\ref{fig:prefix-workflow}, the LLM inference process generally consists of two phases: the prefill phase and the decoding phase. Upon receiving an input sequence, the prefill phase transforms the input tokens into key (K) and value (V) vectors, collectively known as the KV cache. The decoding phase then generates output tokens one by one using this precomputed KV cache.

Specifically, the KV cache is computed by multiplying the hidden states with the K and V projection weights in each Attention module of the transformer layers. As a result, the KV cache is generated layer by layer. When two input sequences share the same prefix, they produce identical KV caches for that shared portion. This enables KV cache reuse, avoiding redundant computation during the expensive prefill phase, especially beneficial for long input sequences. As shown in Figure~\ref{fig:prefix-workflow}, the first request (\textit{[doc1:doc2:query1]}) is processed by the LLM, which generates and stores KV caches for \textit{doc1} and \textit{doc2}. These caches can be stored in HBM, DRAM, or even SSD instead of being discarded. For the second request (\textit{[doc1:doc3:query2]}), which shares the prefix \textit{doc1}, the LLM can reuse the KV cache for \textit{doc1} and only needs to compute the KV cache for \textit{[doc3:query2]}. 
This reuse significantly reduces the computation cost for the second request, especially when \textit{doc1} is long, since the prefill latency grows linearly with input length as stated previously

In long-input scenarios, such as RAG, reusing the prefix KV cache significantly reduces latency during the prefill stage. However, this reuse is restricted to scenarios with identical prefixes because each document's representations become intertwined with preceding documents. For instance, for the third sequence \textit{[doc2:doc3:query3]} in the figure, the LLM cannot reuse KV caches  for \textit{doc2} and \textit{doc3} from previous occurrences due to differing prefixes. Thus, exact prefix matching is essential for KV cache reuse.  While recent research has explored relaxing positional constraints to enable more flexible chunk reuse, these methods typically sacrifice model accuracy~\cite{turborag, agarwal2025cache}. Therefore, we focus solely on prefix-based KV cache reuse to preserve the LLM's accuracy.


What's more, enabling KV-cache reuse with large off-GPU storage (e.g., CPU memory or SSD) introduces data transfer overhead, as cached KVs must be loaded from and written back to these storage devices. Optimizing these data transfers is thus critical to fully realize the performance benefits of KV-cache reuse.

\section{Motivation}

\begin{figure}[t]
    \centering
    \includegraphics[width=0.95\linewidth]{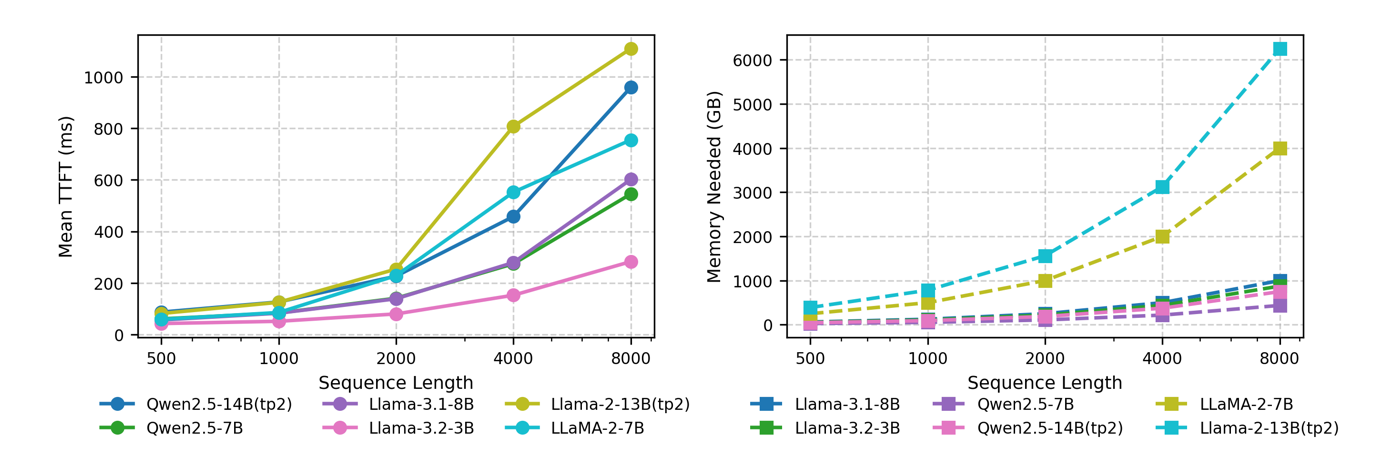}
    \caption{The trend of TTFT and KV cache memory usage as the number of input tokens increases.}
    \Description{Performance comparison plot.}
    \label{fig:kv-stroage-ttft}
\end{figure}

As shown in Figure~\ref{fig:motivation}, to avoid similar performance degradation associated with these naive caching reuse methods, it is necessary to extend SSDs as a fallback medium, further improving the cache system's hit rate. This avoids spending substantial resources on calculating redundant prefilling results. Moreover, blocking synchronous transfers especially between multi-stage are particularly detrimental. By utilizing advanced software and hardware technologies to implement asynchronous RAG workflow control, the overhead of data movement up and down due to the extension of relatively slower SSDs can be hidden behind actual computations.

\begin{figure}[t]
    \centering
    \includegraphics[width=0.95\linewidth]{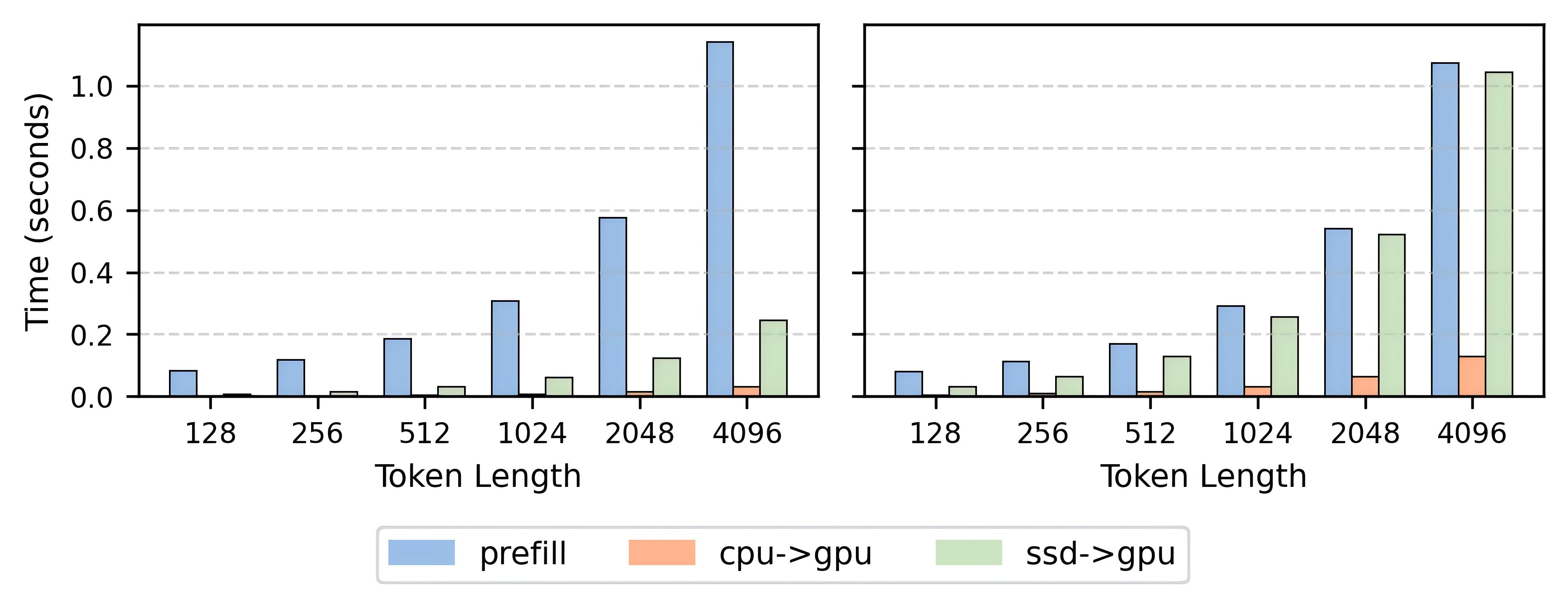}
    \caption{Latency of computation and IO of Qwen2.5-14B (left) and Llama2-13B (right).}
    \Description{Performance comparison plot.}
    \label{fig:compare-io-loading}
\end{figure}

\textbf{Large Storage is Required for KV Cache Reuse.} 
As discussed earlier, KV caches can be reused for identical prefixes. To enable reuse across requests, the generated KV caches must be stored for later access. Traditionally, these caches are kept in GPU memory; however, GPU memory capacity is extremely limited. For example, an NVIDIA H100 GPU with 80 GB of memory can store the KV cache for only about 163,000 tokens of the Llama2-7B model (the smallest in the Llama2 family), excluding model weights and hidden states.
To expand storage capacity, recent works typically offload KV caches to host CPU memory, which offers larger capacity compared to GPU memory. As shown in Figure~\ref{fig:compare-io-loading}, the computation latency is significantly higher than the latency of loading KV caches from CPU memory for both tested models across different token counts. This confirms that reusing KV caches from CPU memory is more efficient than recomputation.
Nevertheless, CPU memory alone remains insufficient for large-scale RAG workloads. As shown in Figure~\ref{fig:kv-stroage-ttft}, when the number of tokens increases to 8192 K, the KV-cache size reaches approximately 0.75 TB for Qwen2.5-14B and 6.23 TB for Llama2-13B, far exceeding the capacity of most host CPU memory, which typically have less than 1 TB size.
In real-world RAG systems, this challenge becomes even more pronounced. For example, the Wikipedia dataset~\cite{wikidataset} contains 8.59 million English documents, totaling around 5 billion tokens. Such scale demands substantially larger storage to sustain high KV-cache reuse efficiency.
As shown in Figure~\ref{fig:compare-io-loading}, the latency of loading KV caches from SSD remains smaller than the recomputation latency in most cases, although its performance is worse than that of CPU memory. This observation suggests that SSDs can serve as a practical external storage layer, enabling additional KV caches to be retained and reused beyond CPU capacity. In our experiments, incorporating a 2 TB SSD improved the cache hit ratio by 10\% compared to using only 256 GB of CPU memory.

\textbf{KV-Cache Transfer Overhead Must Be Optimized.}
When extending KV-cache storage to include CPU memory and SSDs, the system must load the matched KV caches from these devices into GPU memory for computation and offload the newly generated KV caches back to external storage afterward. This process introduces data transfer overhead compared to using GPU memory alone, which can undermine the benefits of KV-cache reuse.
Consider the case of CPU memory. For an input sequence with $N$ tokens, let $N_1$ tokens have matched KV caches and $N_2$ tokens require computation ($N_1 + N_2 = N$). The system must load the KV caches corresponding to the $N_1$ tokens and offload the new KV caches for the $N_2$ tokens after computation. Let $C_1$ denote the time to load (or offload) the KV caches of $N$ tokens (loading and offloading have same cost because the bandwidth between CPU and GPU is same for both directions), and $C_2$ denote the computation time for $N$ tokens. The total processing time $C$ can be expressed as:

\begin{equation}
C = \frac{N_1}{N}C_1 + \frac{N_2}{N}C_2 + \frac{N_2}{N}C_1 = C_1 + \frac{N_2}{N}C_2
\label{eq:load-overhead}
\end{equation}

From this result, we can observe that the data transfer overhead ($C_1$) is constant and independent of the matched token ratio. For example, as shown in Figure~\ref{fig:compare-io-loading}, for the Llama2-13B model with an input of 8k tokens (suppose half are reused), the computation cost is approximately 2 seconds, while the transfer overhead is about 0.5 seconds, resulting in a 25\% performance overhead compared to only considering the computing cost, which cannot be ignored.
When extending storage to SSDs, this overhead becomes even more significant due to the slower read speed of SSDs compared to CPU memory. And even worse, writing data to SSD is usually much slower than read data from it. In our experiments, the tested SSD achieves a read speed of approximately 3 GB/s, but only 500 MB/s for writes. Consequently, synchronous read and write operations to SSDs can, in some cases, result in even worse performance than recomputation.
To preserve the benefits of KV-cache reuse, the transfer overhead must therefore be optimized. Fortunately, the loading, offloading, and computation processes rely on different hardware resources, making it possible to design asynchronous data transfer mechanisms that overlap these operations. Such designs can effectively hide the transfer latency and maximize the performance gains of KV-cache reuse.

In summary, large-scale RAG systems require expanded KV-cache storage across CPU memory and SSDs. At the same time, to offset the increased data transfer cost, asynchronous KV-cache transfer techniques are essential to reduce latency and maintain the efficiency of multi-tier KV-cache reuse.

\section{System Design}

\subsection{System Overview}
\begin{figure}[t]
    \centering
    \includegraphics[width=0.75\linewidth]{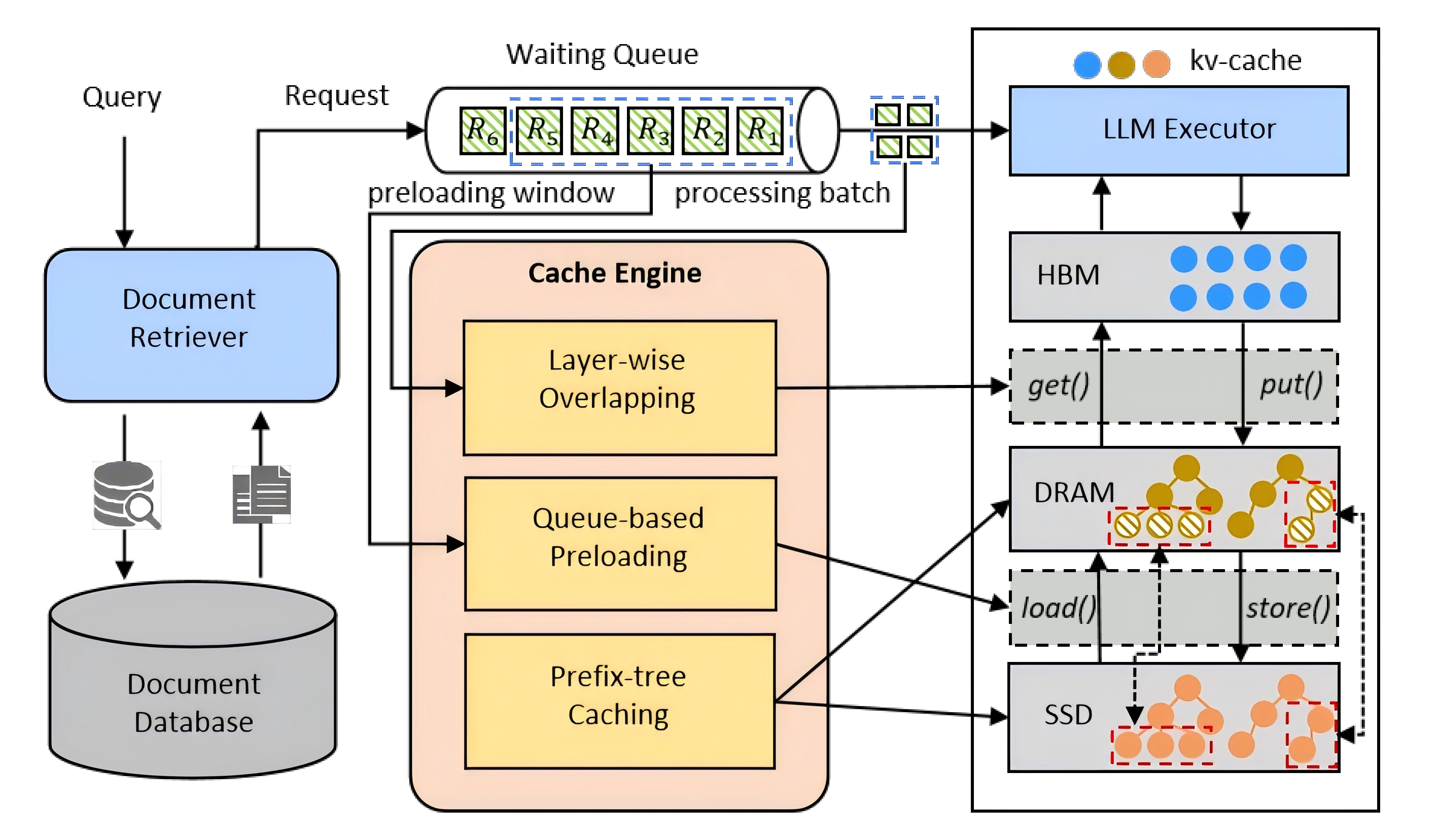}
    \caption{ System overview of \modelname.}
    \Description{System overview plot.}
    \label{fig:system-overview}
\end{figure}

\modelname is a KV-cache management system specifically designed for RAG, aiming to accelerate LLM prefill stage inference by reusing the KV cache of identical documents with prefetch-enhanced techniques. As illustrated in Figure~\ref{fig:system-overview}, \modelname employs a Cache Engine to manage KV-caches stored across DRAM and SSD through three key techniques: (i) a prefix-tree caching scheme that organizes the chunked KV caches with a tree structure, which enables fast prefix matching and improves cache hit ratio with a look-ahead LRU replacement policy. (ii) a layer-wise overlapping technique that leverages the layer-by-layer structure of both LLMs and their KV caches to pipeline loading, computation, and offloading across independent streams, reducing the CPU-GPU communication overhead. (iii) a queue-based prefetching mechanism that prefetches the KV caches from SSD to DRAM when requests are still in waiting queue, thereby hiding SSD loading latency and reducing request waiting time. Equipped with these techniques, \modelname extends KV-cache storage capacity beyond GPU memory to DRAM and SSD, while simultaneously reducing the transfer overhead associated with loading and offloading operations, maximizing the KV-cache reuse benefits.

When a user submits a query, the document retriever first searches the document database to identify relevant documents, combines them with the query, and sends the resulting request to a waiting queue for processing by the LLM executor. During execution, the LLM executor interacts with the Cache Engine to identify reusable prefixes by traversing the prefix tree of stored KV caches. It then performs layer-wise overlapping, loading existing KV caches from DRAM while simultaneously computing and offloading new KV caches layer by layer, thus minimizing CPU–GPU data transfer latency. Meanwhile, the Cache Engine proactively monitors requests in the waiting queue, updating the priorities of KV-cache chunks that will be used soon to prevent their eviction, and prefetching corresponding KV caches from SSD into DRAM ahead of execution. Through this coordinated process, \modelname effectively reduces KV-cache transfer overhead, fully exploits prefix-based reuse opportunities, and significantly enhances overall inference performance.


\subsection{Prefix-Tree Caching}

\begin{figure}[t]
    \centering
    \includegraphics[width=0.75\linewidth]{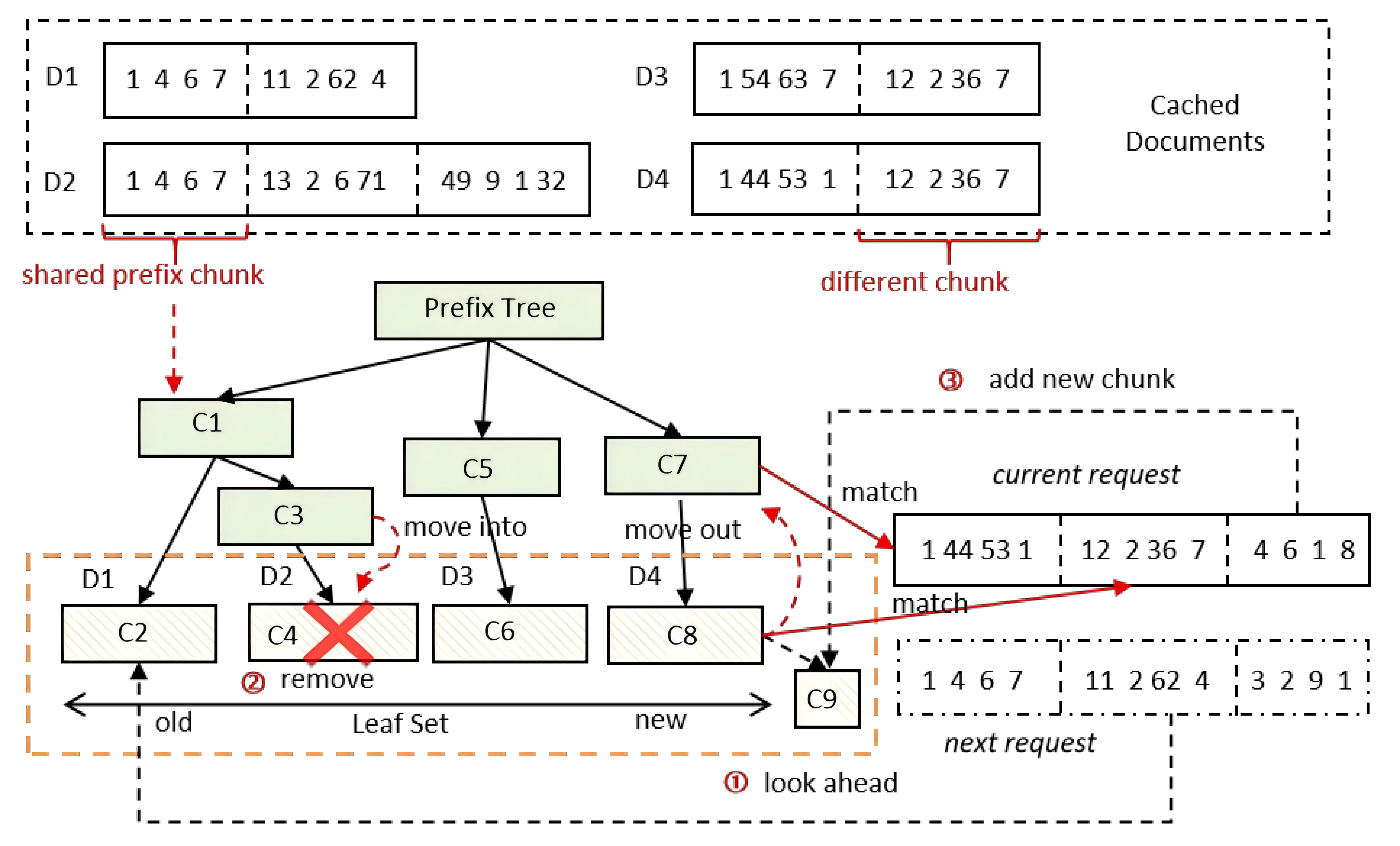}
    \caption{Prefix-tree KV cache.}
    \label{fig:prefixcache}
    \Description{Prefix tree method plot.}
\end{figure}

To efficiently manage the KV cache across CPU memory and SSD, an effective organization strategy is required. Since our focus is on prefix KV-cache reuse, we adopt a prefix tree structure to manage the cache.

As illustrated in Figure~\ref{fig:prefixcache}, each document is divided into fixed-size chunks to facilitate efficient management, and these chunks are organized in a prefix-tree structure. Chunks sharing identical prefixes are mapped to the same tree node. For example, the first chunks of documents \textit{D1} and \textit{D2} are identical, and thus both are associated with node \textit{C1}. However, since the KV cache is position-dependent, only chunks with identical prefixes can be reused. Even if the second chunks of \textit{D1} and \textit{D2} contain the same token IDs, their differing prefixes lead to distinct cache nodes (\textit{C6} and \textit{C8}, respectively). During query processing, each incoming request is matched against the tree in a chunk-wise manner, starting from the root node, until a mismatch occurs. In the example shown, the current request matches two chunks (\textit{C7} and \textit{C8}), implying that only the remaining unmatched chunks must be newly computed.

When the cache reaches its capacity, certain chunks need to be evicted to accommodate new ones. Because each chunk depends on its parent node and can only be reused when its parent also resides in the cache, eviction is restricted to the leaf nodes of the tree. As illustrated in the figure, all leaves are maintained using an LRU (Least Recently Used) cache replacement policy, and eviction decisions are made based on recency. To further improve the cache hit ratio, we introduce a look-ahead update mechanism that leverages pending requests from the waiting queue to predict near-future reuse and protect corresponding chunks from eviction. For instance, when one chunk must be evicted to insert a new chunk \textit{C9}, a look-ahead to the next request increases the priority of \textit{C2} (the oldest chunk). Consequently, instead of evicting \textit{C2}, the algorithm removes the second-oldest leaf, \textit{C4}, and inserts \textit{C9} as a child of \textit{C8}. This decision preserves \textit{C2} for imminent reuse, thereby enhancing the overall hit ratio. Furthermore, when \textit{C4} is evicted, its parent becomes a new leaf and is added to the leaf set; similarly, upon inserting \textit{C9}, \textit{C8} transitions from a leaf to an internal node and is removed from the leaf set. After this update, the leaf set changes from \textit{\{C2, C4, C6, C8\}} to \textit{\{C6, C2, C3, C9\}}, in contrast to the \textit{\{C4, C6, C9\}} produced under the conventional LRU policy.

\begin{figure}[t]
    \centering
    \includegraphics[width=0.75\linewidth]{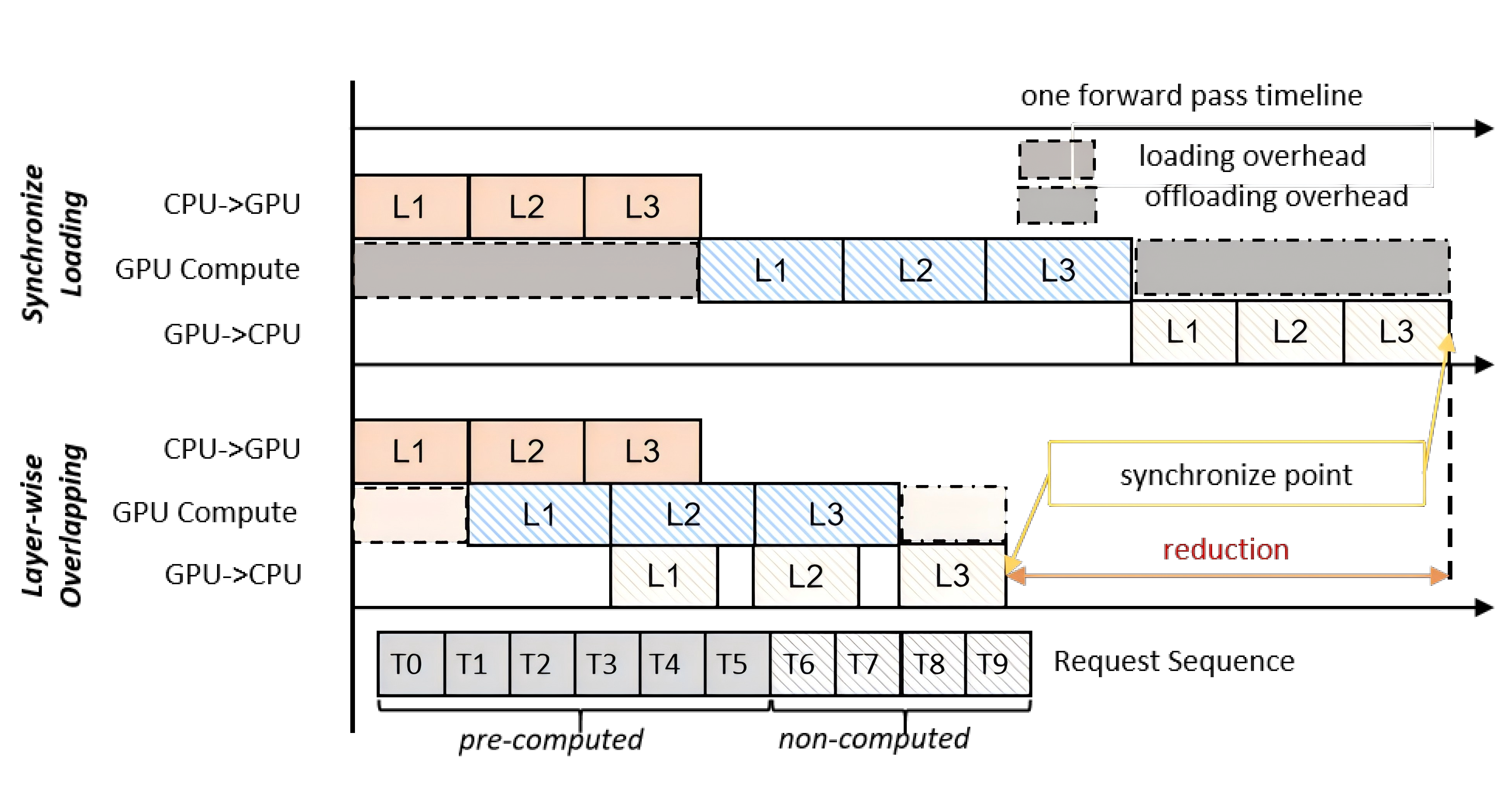}
    \caption{Layer-wise overlapping the compute and transfer.}
    \label{fig:layer-wise}
    \Description{Layer overlap plot.}
\end{figure}

By integrating the prefix-tree structure with the look-ahead LRU policy, our system enables efficient KV-cache management for prefix-based matching and achieves a higher cache hit ratio.

\begin{figure}[t]
    \centering
    \includegraphics[width=0.8\linewidth]{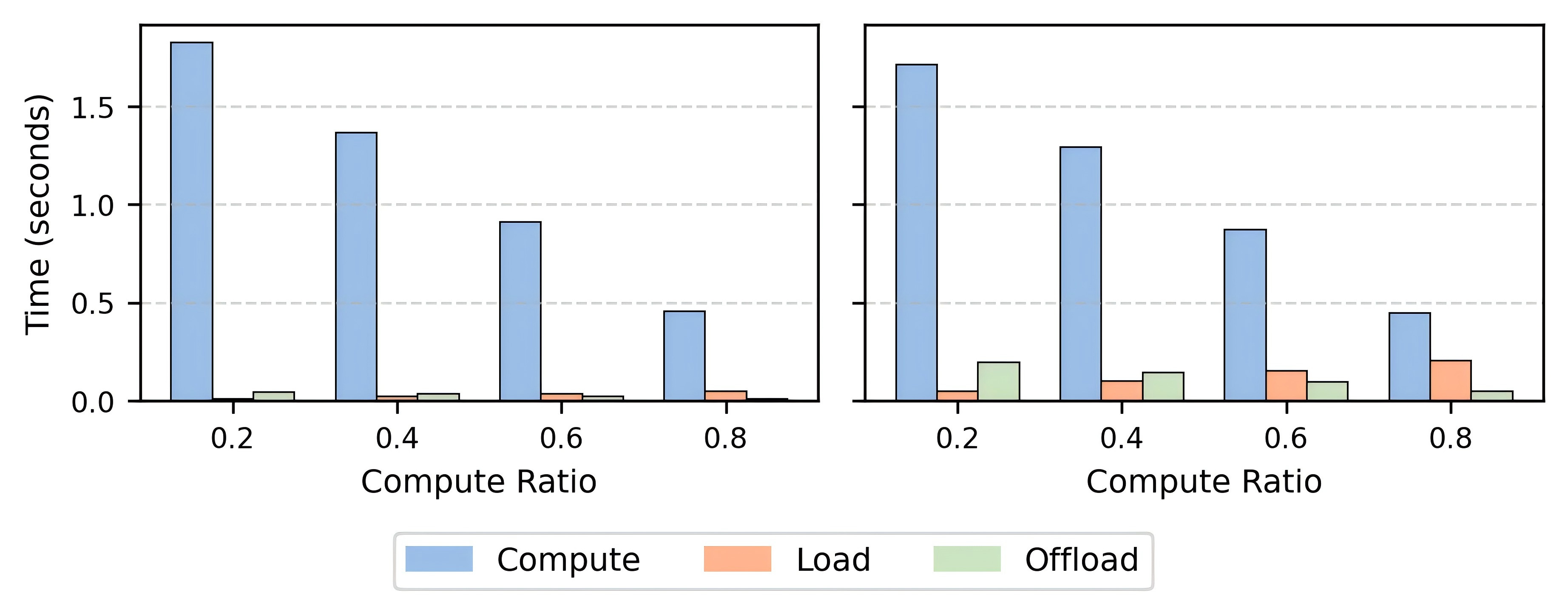}
    \caption{Compare under different computed ratio of Qwen2.5-14B (left) and Llama2-13B (right).}
    \label{fig:diff-rate-computed}
    \Description{Performance comparison plot.}
\end{figure}

\subsection{Layer-wise Overlapping}

In the KV-cache reuse pipeline, the system must load precomputed KV-cache entries from the CPU to the GPU, process the remaining non-cached tokens using the LLM, and subsequently store the newly generated KV cache back from the GPU to the CPU. As illustrated in Figure~\ref{fig:layer-wise}, these load and offload operations can introduce significant overhead during a single forward pass due to transfer latency between the CPU and GPU when executed synchronously.

Given the layer-wise structure of LLMs, the KV cache is naturally organized in a layer-by-layer manner. This structural characteristic enables layer-wise overlapping, where computation can begin as soon as the KV cache of the first layer is loaded without waiting for the complete cache transfer. Similarly, offloading can commence immediately after each layer’s computation finishes, rather than waiting until the entire forward pass is complete. This overlapping approach effectively hides most of the data transfer latency, limiting the cost to approximately one layer’s transfer time. Compared with the synchronous transfer overhead $C_1$ defined in Equation~\ref{eq:load-overhead}, the layer-wise overlapping theoretically reduces the overhead to $\frac{1}{n}C_1 = \frac{N_1}{N}C_1 \times \frac{1}{n} + \frac{N_2}{N}C_1 \times \frac{1}{n}$, where $n$ denotes the number of layers in the LLM, which typically exceeds 32 for models with 7B parameters or more, thus achieving a substantial reduction in transfer overhead.

To ensure effective overlap, the load/offload latency for each layer must be smaller than the corresponding computation latency.
For offloading, the overhead grows linearly with computation since the newly generated KV cache size is proportional to the number of processed tokens. Consequently, as shown in Figure~\ref{fig:compare-io-loading}, the offloading latency remains lower than the computation latency for the same token count.
For loading, a larger number of precomputed tokens increases transfer latency while reducing computation latency for a fixed sequence length. Nevertheless, due to high transfer bandwidth, the loading latency remains below the computation time, even when precomputed tokens dominate the sequence. As shown in Figure~\ref{fig:diff-rate-computed}, for the Qwen2.5-14B model with an 8192-token context, even when the computed ratio reaches 80\%, the loading latency is still shorter than computation time. Therefore, the layer-wise overlap effectively hides both loading and offloading costs.
Moreover, in modern efficient inference systems~\cite{kwon2023efficient, zheng2024sglang}, the prefill and decoding stages are often fused into a single forward pass. This integration further increases computation time, thereby enhancing the opportunity to overlap KV-cache transfers with computation.

In summary, the proposed layer-wise overlapping technique substantially mitigates KV-cache transfer overhead between CPU and GPU, maximizes the benefits of KV-cache reuse, and accelerates the prefill stage.

\subsection{Queue-base Prefetch}

\begin{figure}[t]
    \centering
    \includegraphics[width=0.75\linewidth]{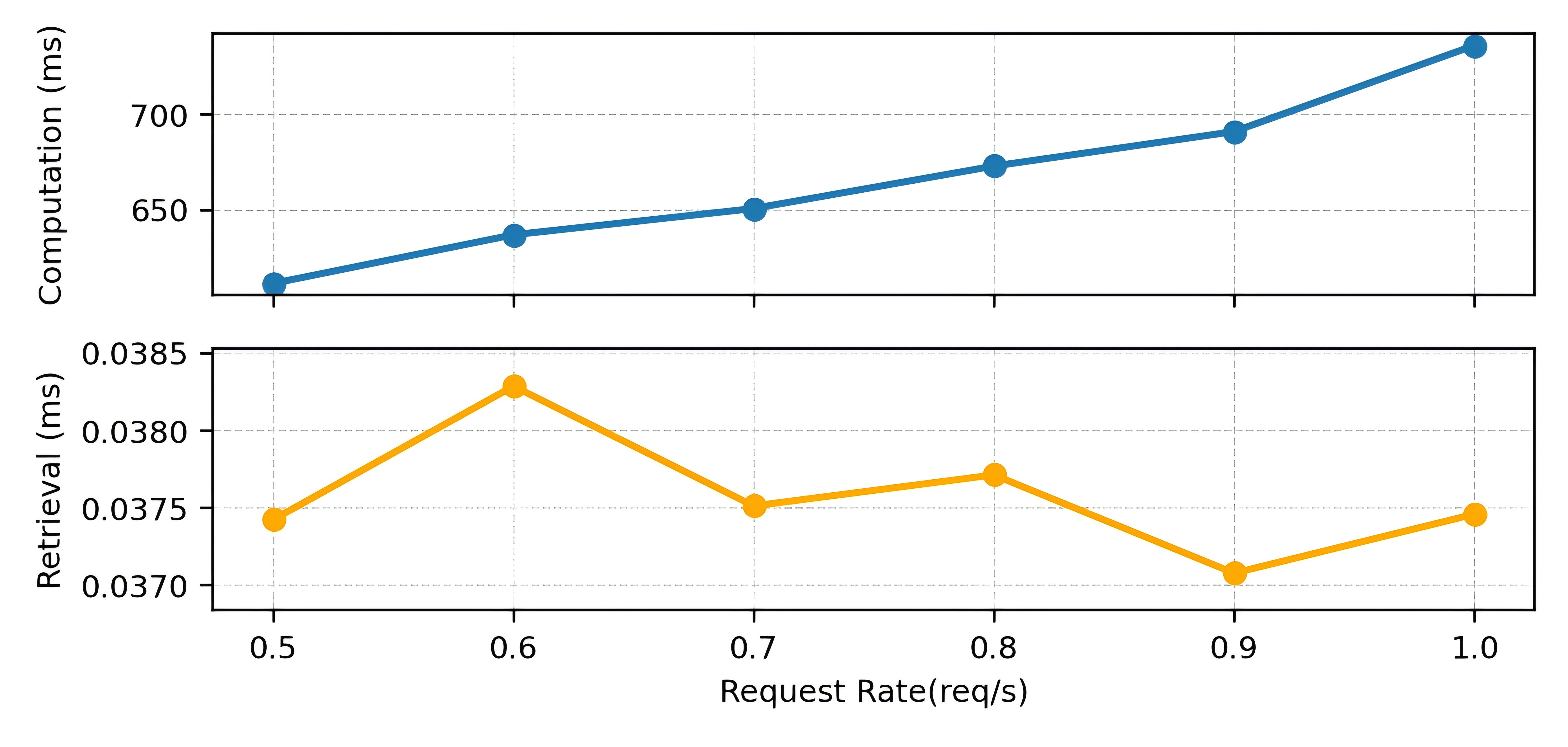}
    \caption{Latency of retrieval and generation.}
    \label{fig:retrivel-compuation}
    \Description{Performance comparison plot.}
\end{figure}


\begin{figure}[t]
    \centering
    \includegraphics[width=0.95\linewidth]{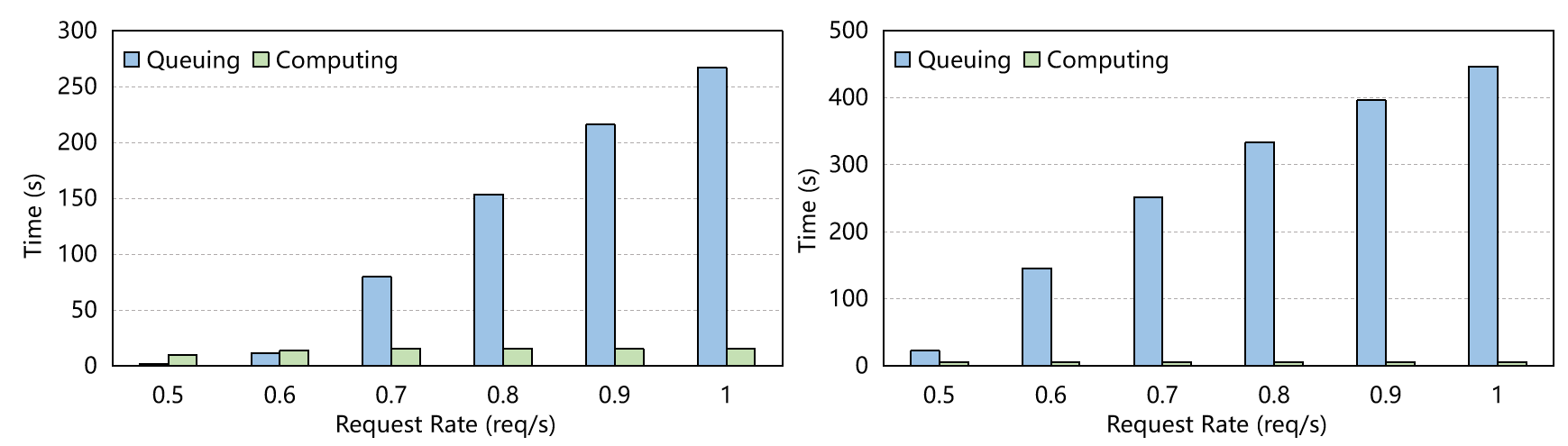}
    \caption{Comparison of Queuing and Computing for Qwen2.5-14B (left) and Llama2-13B (right).}
    \label{fig:queue-time}
    \Description{Performance comparison plot.}
\end{figure}

\begin{figure}[t]
    \centering
    \includegraphics[width=0.75\linewidth]{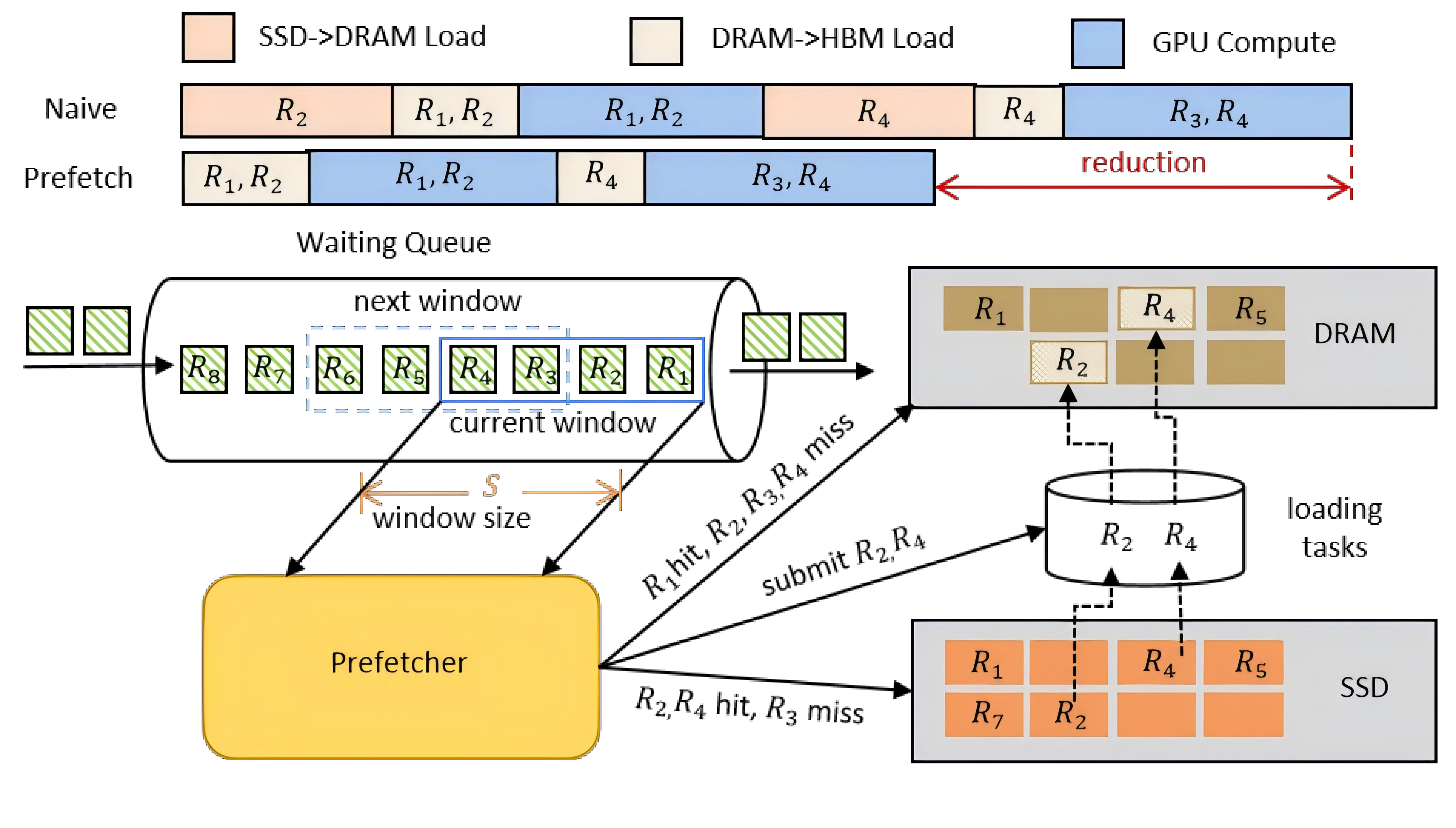}
    \caption{Queue-based prefetching the kv-cache from SSD to DRAM.}
    \label{fig:prefetch}
    \Description{Prefetch plot.}
\end{figure}

Due to the significantly slower read speed of SSD, on-demand loading of KV caches from SSD to GPU introduces substantial waiting overhead. Unlike CPU–GPU data transfers, this process cannot benefit from layer-wise overlapping, since the SSD loading latency often exceeds the computation latency when the proportion of precomputed tokens is large. Therefore, additional techniques are required to mitigate this bottleneck.

Fortunately, as shown in Figure~\ref{fig:retrivel-compuation}, the document retrieval process is considerably faster than model inference across different request rate for both evaluated models. This observation implies that requests in the waiting queue have already completed their document retrieval phase, creating an opportunity to prefetch their KV caches from SSD into DRAM before computation begins. Moreover, Figure~\ref{fig:queue-time} shows that under heavy workloads, requests typically experience long queuing compared to computing, which can be effectively utilized for KV-cache prefetching.

As illustrated in Figure~\ref{fig:prefetch}, the prefetcher continuously monitors the waiting queue and proactively loads KV caches for upcoming requests. Specifically, the prefetcher maintains a look-ahead window to identify requests within a certain range of the queue. It then checks the KV-cache status of these requests in both DRAM and SSD. If a KV cache is found on the SSD but not yet in DRAM, the prefetcher launches asynchronous loading tasks to transfer it into DRAM. For example, as shown in the figure, the prefetcher detects that the KV caches corresponding to requests $R_2$ and $R_4$ reside on the SSD but are absent from DRAM, thus initiating their transfer. In contrast, request $R_1$ already has its cache in DRAM, requiring no further action, while request $R_3$ misses in both DRAM and SSD, necessitating a full recomputation of its tokens. Compared with a naïve on-demand strategy, this prefetching mechanism effectively eliminates the SSD-to-DRAM loading delay for most requests, significantly accelerating overall processing. After each prefetching cycle, the look-ahead window slides forward to track subsequent requests.

Furthermore, storing newly generated KV caches is also performed asynchronously. As previously discussed, after completing one forward pass, the system must write all KV caches back into CPU memory. Once this step is done, the Cache Engine immediately submits asynchronous write-back tasks to persist the data onto SSDs, allowing the system to begin the next iteration without waiting for the disk write operations to finish.

In summary, the system employs a prefetching mechanism to proactively fetch KV caches for requests still waiting in the queue, thereby mitigating the SSD-to-DRAM loading overhead. Combined with asynchronous write-back, this approach effectively leverages SSD-based cache extensions while maintaining low latency.

\section{Implementation}
\begin{figure}[t]
    \centering
    \includegraphics[width=0.95\linewidth]{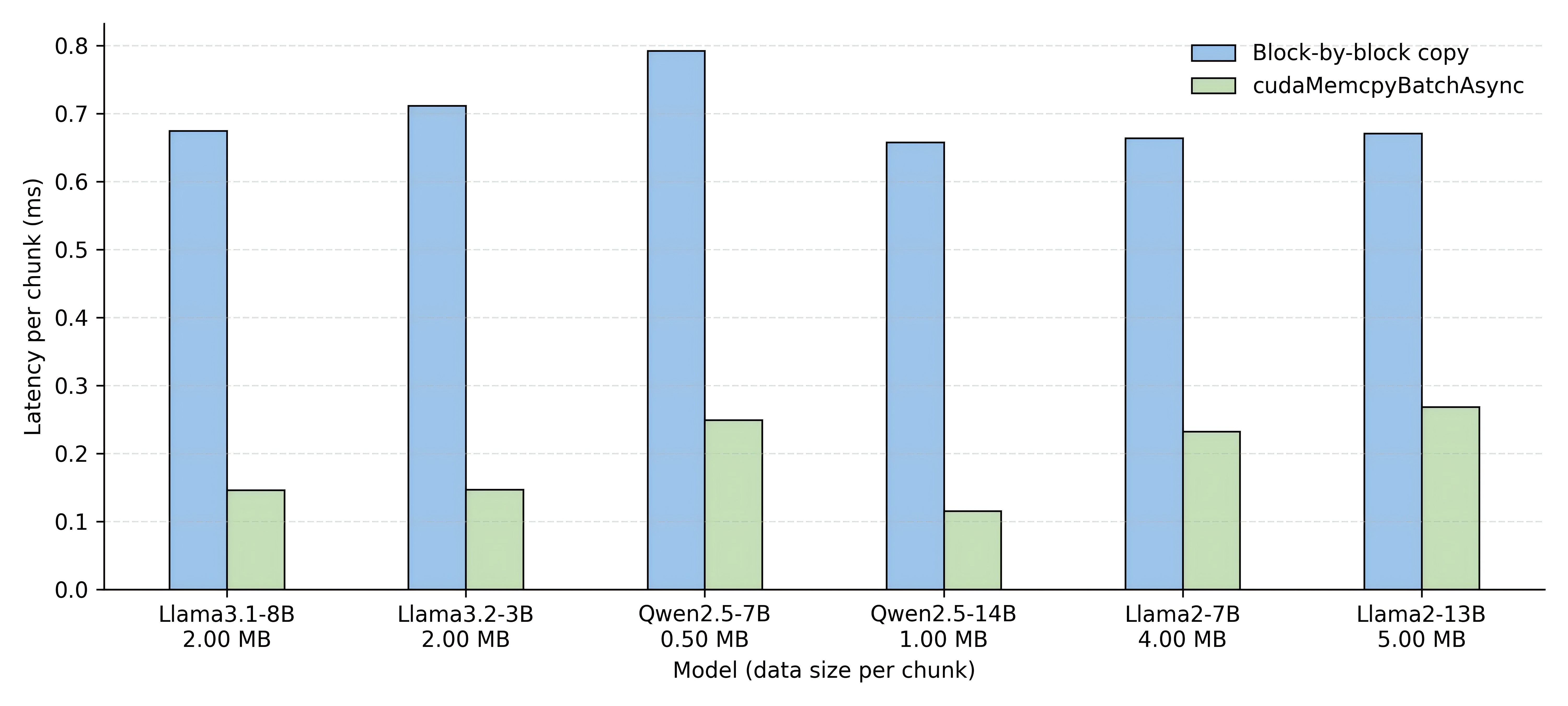}
    \caption{Chunk KV Cache transfer performance of block by block approach and BatchAsync.}
    \label{fig:api}
    \Description{Api comparision plot.}
\end{figure}
We developed \modelname upon vLLM~\cite{kwon2023efficient}, a state-of-the-art LLM serving system. vLLM employs the \textit{PagedAttention} technique to efficiently manage the KV cache in GPU memory. This technique partitions each sequence's KV cache into small blocks, akin to virtual memory management in operating systems. PagedAttention allows mapping several non-consecutive small physical blocks to a large, virtually consecutive KV cache block for a sequence, thereby maximizing memory utilization. For real implementation, vLLM consists of two main components, scheduler and worker. The scheduler maintains incoming request waiting and running queues, manages KV cache blocks via a block table, and assigned KV cache blocks for processing requests. The workers execute the LLM to process inputs using the KV cache blocks assigned by the scheduler. What's more, the KV cache is allocated layer by layer, rather than allocate a whole block for all layers, so we can do layer-wise memory copying. 

As previously stated, we split sequences into chunks and manage their KV cache chunks in CPU memory and SSD by constructing a prefix tree structure, leaving GPU memory management to vLLM. For efficient management and movement, our chunk size is typically larger than the block size of vLLM (256 vs. 16 in the our setting). To efficiently copy KV cache from a CPU chunk to multiple non-consecutive GPU memory blocks, we leverage the new CUDA API \textit{cudaMemcpyBatchAsync()}, supported by CUDA 12.8 or higher. This API performs a single, batch copy for all blocks, significantly reducing kernel launch overhead compared to block-by-block copying with \textit{cudaMemcpyAsync()}. For instance, copying a chunk of one layer KV cache of Llama2-13B, takes 0.671 ms with block-by-block copying, whereas \textit{cudaMemcpyBatchAsync()} achieves this in only 0.261 ms on a 32GB/s PCIe connection, as Figure~\ref{fig:api}. To enable layer-wise overlapping, we create three distinct CUDA streams for CPU-to-GPU transfer, GPU computation, and GPU-to-CPU transfer, allowing these operations to execute in parallel. For queue-based prefetching and look-ahead LRU cache policy, we monitor the request waiting queue within the vLLM scheduler. We send the waiting requests within a preloading window (set to 4 in our configuration) to the cache engine of \modelnamenospace,  and update the recency for matched chunks and use  a dedicated thread (e.g., Prefetcher) to fetch the KV cache from SSD to CPU memory upon prefix matching.

\begin{algorithm}
\caption{PCR\_step (Simplified)}\label{alg:pcr_step}
\begin{algorithmic}[1]
\Require $\textit{scheduler\_output}$, $\textit{cached\_reqs}$, $\textit{kv\_caches}$

\Statex \Comment{--- Prefetch guided by scheduler ---}
\ForAll{$\textit{req} \in \Call{Reverse}{\textit{scheduler\_output.prefetch\_reqs}}$}
    \ForAll{$\textit{chunk} \in \Call{Chunkify}{\textit{req}, \textit{CHUNK\_SIZE}}$}
        \State $\textit{key} \gets \Call{HashPrefix}{\textit{chunk}}$
        \If{in CPU cache} \Call{BumpPriority}{\textit{key}}
        \ElsIf{in SSD cache} \Call{SubmitSSDToCPULoad}{\textit{key}}
        \Else \textbf{break} \EndIf
    \EndFor
\EndFor
\Call{EnforcePrefetchWindow}{}

\Statex \Comment{--- Plan data movement for scheduled requests ---}
\ForAll{$\textit{req} \in \Call{ScheduledRequests}{\textit{scheduler\_output}, \textit{cached\_reqs}}$}
    \ForAll{$(\textit{chunk}, \textit{parent}) \in \Call{PrefixChunks}{\textit{req}}$}
        \State $\textit{key} \gets \Call{HashPrefix}{\textit{chunk}, \textit{parent}}$
        \If{in CPU} add to $\textit{cpu\_to\_gpu}$
        \ElsIf{in SSD} trigger load and add to $\textit{ssd\_to\_gpu}$
        \Else add to $\textit{gpu\_to\_cpu}$ \EndIf
    \EndFor
    \Call{AdjustTokens}{\textit{req}}
\EndFor

\Statex \Comment{--- Execute async transfers and finalize ---}
\If{$\textit{cpu\_to\_gpu} \neq \emptyset$} \Call{LayerwiseUpload}{\textit{cpu\_to\_gpu}} \EndIf
\If{$\textit{gpu\_to\_cpu} \neq \emptyset$} \Call{LayerwiseDownload}{\textit{gpu\_to\_cpu}} \EndIf
\Call{DrainCompletedSSDLoads}{}
\Call{WaitForUploadEvents}{\textit{kv\_caches}}
\Call{ShrinkPrefetchWindow}{}
\end{algorithmic}
\end{algorithm}

The main workflow of \modelname is illustrated in Algorithm~\ref{alg:pcr_step}. It implements a coordinated execution pipeline that tightly integrates caching, multi-tier storage management, and compute scheduling. At its core, it uses scheduler-provided hints to drive prefetching within a bounded window, preventing overfetching; tracks data across three tiers consist of CPU, SSD, and GPU to maintain hot chunks in fast memory; performs layerwise batched memory copies on dedicated streams to amortize per-call overhead and hide PCIe stalls; and enforces strict coupling between the scheduler and data availability, ensuring computation proceeds only when required chunks are confirmed present. The design emphasizes end-to-end coordination and pipelining to jointly optimize throughput and latency.




\section{Evaluation}


\subsection{Experimental Methodology}

\noindent \textbf{Hardware.}
We evaluate \modelname on two distinct hardware platforms to assess its performance under different system configurations. The first system is equipped with two NVIDIA A6000 GPUs (each with 48 GB of memory), 256 GB of host CPU memory, and 96 CPU cores, along with a 4 TB NVMe SSD for storage. The second system features two NVIDIA RTX 4090 GPUs (each with 24 GB of memory), 128 GB of host CPU memory, 128 CPU cores, and an identical 4 TB NVMe SSD.

Both systems use PCIe 4.0 to connect the CPUs and GPUs, which provides a theoretical bidirectional bandwidth of up to 32 GB/s per GPU (approximately 24 GB/s in real-world measurements due to protocol overhead and system contention). The SSD delivers sequential read and write speeds of around 3 GB/s, ensuring minimal I/O bottleneck during data loading. This setup allows us to study the impact of GPU memory capacity, CPU resources, and system architecture on \modelname’s inference latency and throughput across realistic deployment environments.

\noindent \textbf{Models.}
To assess the performance of \modelname across various model architectures, we use two popular LLM families: Qwen and Llama. Specifically, we evaluate six models: Llama3.1-8B, Llama3.2-3B, Qwen2.5-8B, Qwen2.5-14B, Llama2-7B, and Llama2-13B. The primary distinction between Qwen2.5 and Llama2 in terms of KV cache is that Llama3 and Qwen2.5 uses GQA, whereas Llama2 employs MHA. GQA generally has fewer KV heads than MHA, resulting in a smaller KV cache size. Due to memory constraints, Qwen2.5-14B and Llama2-13B are executed on two GPUs, whereas all other models fit comfortably within the memory of a single GPU.

\noindent \textbf{Workloads.}
For simulating the RAG workflow, we use the Wikipedia dataset as the corpus for retrieval and the SQuAD dataset for queries. Documents and queries are embedded using the MiniLM embedding model, with the two most relevant documents retrieved for each query. We first construct two datasets consisting of 1,000 and 2000 inputs, each input containing two documents and one query. The average length of the workload is approximately 6.8k tokens, with a 40\% and 35\% repetition ratio for KV cache reuse respectively. To control the test time, we perform 2,000 sampling iterations on both datasets. For the vLLM implementation, dataset 1 consists of full sampling followed by 1,000 rounds of oversampling with replacement, while dataset 2 uses full sampling without oversampling. Since we focus on prefill latency during generation, we set the output length to 16 tokens for all tests. Following previous works, request arrival times are modeled using a Poisson process parameterized by the arrival rate.

\noindent \textbf{Baselines.}
We compare \modelname against two existing LLM serving systems and two simplified implement of our system to evaluate its efficiency: (1) vLLM~\cite{kwon2023efficient}, A well-known system that uses PagedAttention to mitigate memory fragmentation and block-level prefix caching to enable KV-cache reuse in GPU memory. (2) LMCache~\cite{liu2025LMCache}, A state-of-the-art KV cache prefix reuse and prefetching system built on vLLM’s KV cache manager via its connector mechanism. It delivers high performance by leveraging prefix sharing across the GPU–CPU–SSD memory hierarchy. (3) CCache, An extension of vLLM that utilizes CPU memory to store additional KV cache for reuse. (4) SCCache, A further extension of CCache, incorporating SSD storage to expand KV cache capacity. All evaluated methods share vLLM as their common backbone, ensuring consistency in other integrated optimizations and guaranteeing a fair comparison in our experiments. 

\subsection{Overall Performance}

\begin{figure}[t]
    \centering
    \includegraphics[width=0.85\linewidth]{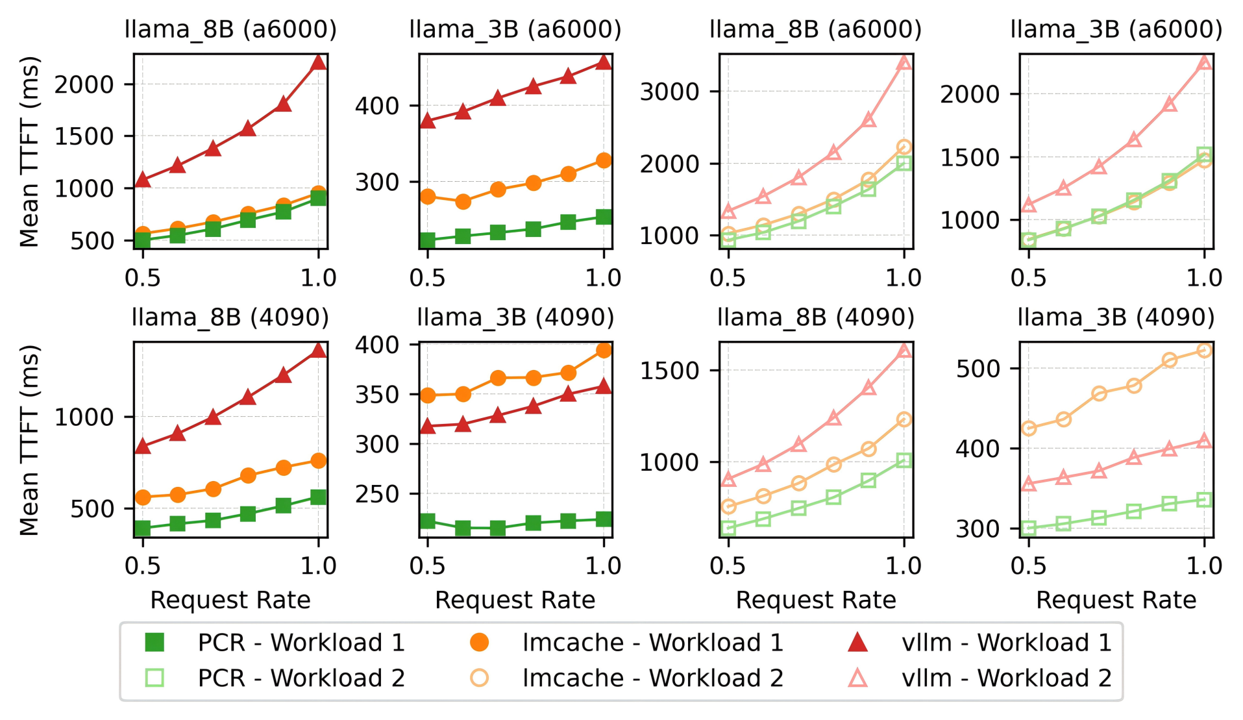}
    \caption{TTFT of two models across hardware platforms and workload configurations.}
    \label{fig:ttft_all}
    \Description{Line plots showing mean TTFT for Llama-8B and Llama-3B on A6000 and RTX 4090 GPUs under two sampling configurations (config1 and config2), with three caching methods compared.}
\end{figure}


To evaluate the performance of \modelnamenospace, we conduct the experiments described before and obtain the results shown in Figure~\ref{fig:ttft_all}. Across all experimental scenarios, including two different hardware platforms, multiple model sizes, request rates ranging from 0.5 to 1.0, and varying input sampling strategies—our method consistently achieves the fastest TTFT (Time To First Token). In most cases, LMCache also outperforms the native vLLM framework. In general, TTFT increases as the request rate rises; however, our system exhibits a significantly flatter growth curve compared to both baselines, indicating that its advantage becomes more pronounced under higher workload. For instance, when running the Llama-8B model on an RTX 4090 GPU, \modelnamenospace achieves speedups of 2.13× and 1.42× over the vLLM baseline under Workload 1 and Workload 2, respectively, improving further to 2.47× and 1.59× at higher request rates. Correspondingly, the absolute latency reduction grows from 445 ms and 267 ms to 797 ms and 598 ms, highlighting \modelnamenospace’s superior scalability under intensive workloads.

\begin{figure}[t]
    \centering
    \includegraphics[width=0.85\linewidth]{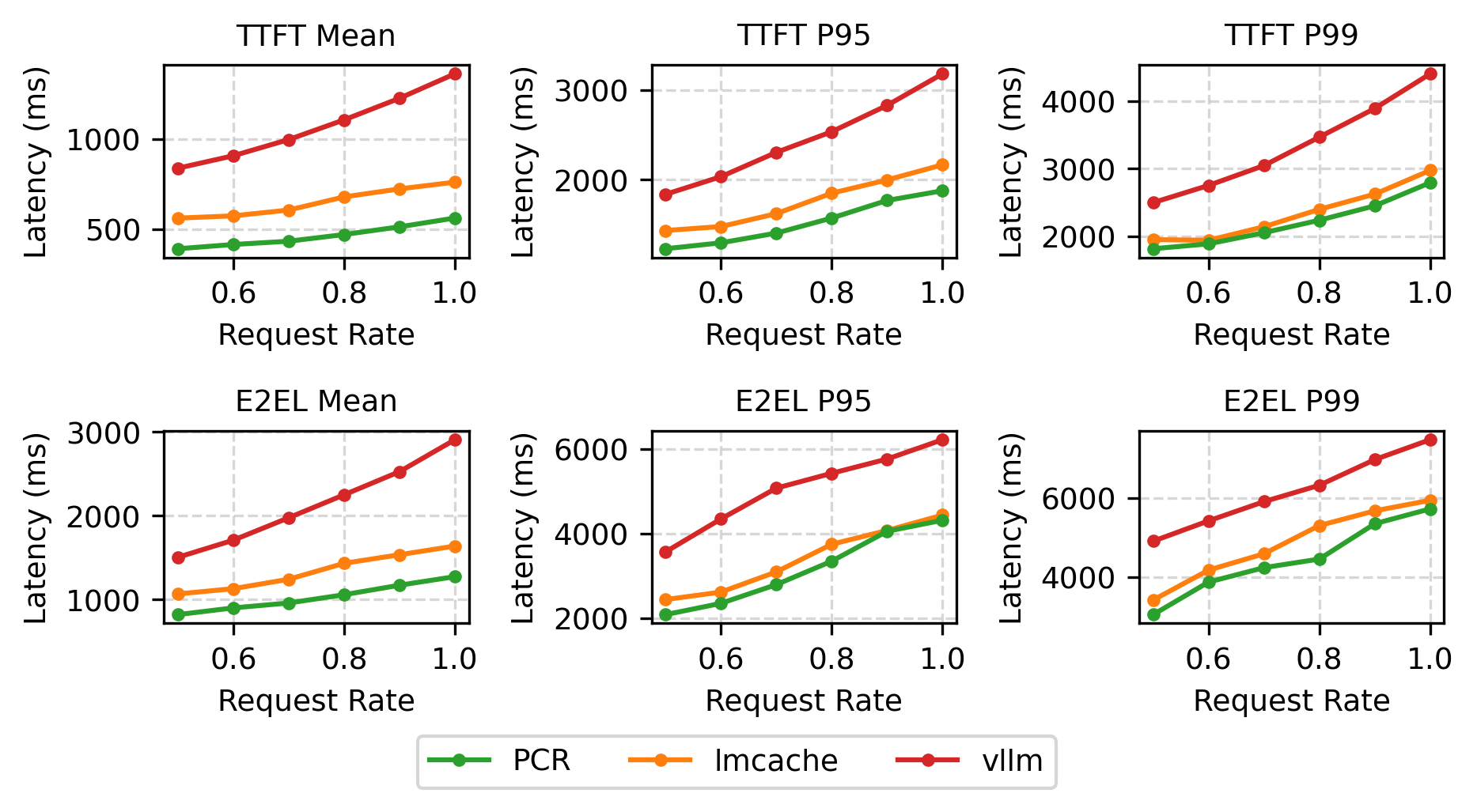}
    \caption{TTFT and E2EL metrics of Llama-8B.}
    \label{fig:tail_distribution}
    \Description{TTFT and E2EL metrics of Llama-8B.}
\end{figure}

As Figure~\ref{fig:tail_distribution} shows, our system maintains consistent superiority not only in mean latency but also across key tail-latency metrics. For example, at a high request rate of 0.9, PCR achieves a P99 end-to-end latency (E2EL) of just 86 ms, compared to 124 ms for LMCache and 142 ms for vLLM—a reduction of over 30\%. Similarly, in TTFT P95, PCR delivers 58 ms, while LMCache and vLLM incur 89 ms and 103 ms, respectively.
Critically, real-world service quality depends heavily on these tail percentiles: they reflect the experience of nearly all users, not just the average case. The fact that PCR outperforms competitors across all six metrics—TTFT and E2EL at mean, P95, and P99—demonstrates that its efficiency gains are both broad and robust. In practice, this means almost every user request benefits from lower latency and more predictable responsiveness, especially under load.

\begin{figure}[t]
    \centering
    \includegraphics[width=0.75\linewidth]{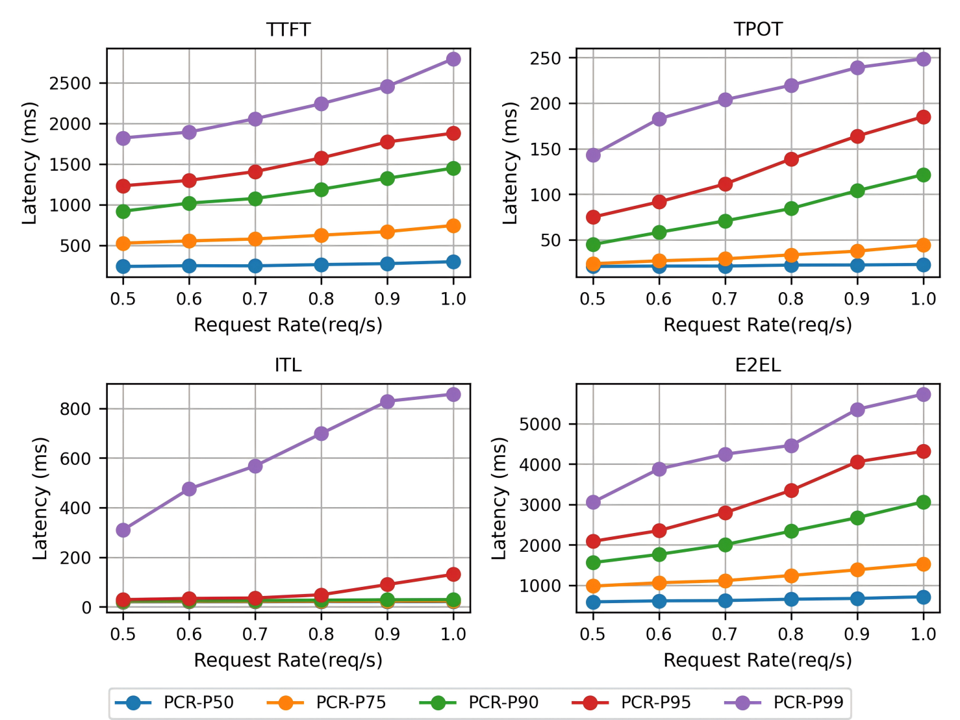}
    \caption{Typical percentile scalability results for various latency metrics.}
    \label{fig:distribution}
    \Description{p metrics of Llama-8B.}
\end{figure}

The evaluation results in Figure~\ref{fig:distribution} demonstrate that \modelnamenospace achieves strong latency stability under increasing load. As shown in Figure, all latency metrics exhibit smooth, monotonic growth across request rates from 0.5 to 1.0 req/s, with no abrupt spikes or saturation points, indicating robust system behavior. Notably, the median (P50) latencies remain consistently low, TTFT stays below 300 ms and E2EL under 600 ms even at peak load. Crucially, tail latencies are well-controlled, ITL P99 increases only from ~300 ms to ~850 ms, and E2EL P99 rises gradually from ~3,000 ms to ~6,000 ms, reflecting effective mitigation of jitter and stalls. The narrow gap between P75 and P90 across all metrics further confirms low variability in the majority of requests. Moreover, the moderate slope of P99 curves suggests \modelnamenospace’s scheduling and caching mechanisms successfully prevent worst-case degradation, ensuring predictable performance. Together, these characteristics highlight our method's ability to deliver both low-latency responsiveness for typical requests and reliable tail behavior under pressure, making it suitable for production environments demanding consistent quality of service.

\subsection{Performance Breakdown}

\begin{table}[t]
\centering
\caption{Performance breakdown for \modelnamenospace.}
\small
\setlength{\tabcolsep}{4pt}
\begin{tabular}{c|c|cc|cc}
\toprule
 & \multirow{2}{*}{\textbf{Techniques}} & \multicolumn{2}{c|}{\textbf{0.5 request/s}} & \multicolumn{2}{c}{\textbf{1.0 request/s}} \\
 & & \textbf{Time To First Token } & \textbf{Reduction} & \textbf{Time To First Token} & \textbf{Reduction} \\
\midrule
\multirow{4}{*}{\textbf{Qwen2.5-7B}} 
 & base         &0.778 s   & -  & 1.282 s  & - \\
 & +overlap   & 0.762 s & 2.05\% & 1.217 s & 5.07\% \\
 & +prefetch    & 0.765 s & 1.67\% & 1.207 s & 5.85\% \\
\midrule
\multirow{4}{*}{\textbf{Qwen2.5-14B}} 
 & base         & 2.716 s  & -  & 139.486 s   & -  \\
 & +overlap   & 2.586 s & 4.78\% & 124.963 s & 10.41\% \\
 & +prefetch    & 2.583 s &  4.89\% & 122.678 s & 12.05\% \\
 \midrule
\multirow{4}{*}{\textbf{Llama2-7B}} 
 & base         & 1.634 s   & -  & 11.133 s  & - \\
 & +overlap   & 1.344 s & 17.74\% & 5.841 s & 47.53\% \\
 & +prefetch    & 1.293 s & 20.86\% & 3.441 s & 69.09\% \\
 \midrule
\multirow{4}{*}{\textbf{Llama2-13B}} 
 & base         & 15.132 s   & -  & 487.427 s  & -  \\
 & +overlap   & 9.490 s & 37.28\% & 431.354 s & 11.50\% \\
 & +prefetch    & 6.191 s & 59.08\% & 335.573 s & 31.15\% \\
\bottomrule
\end{tabular}
\label{tab:performance-breakdown}
\end{table}





To evaluate the contribution of layer-wise overlapping and queue-based prefetching independently, we perform a performance breakdown under both high and low request rates. As shown in Table~\ref{tab:performance-breakdown}, both techniques provide measurable benefits. Among them, layer-wise overlapping yields the largest reduction in latency (averaging 15\%), as the offloading process is particularly costly since all newly generated KV caches must be transferred while only one fraction KV cache will be reused. What's more, two Llama models gain larger benefits from prefetching because its larger KV cache size, which cause more KV cache loading from SSD compared to Qwen models under same CPU memory budget. Specially, under the experiment setting, the Qwen models only gain about 10\% KV cache reuse in SSD while Llama models gain 20\% KV cache hit rate from SSD.  Furthermore,  prefetching technique achieve greater improvements under high request rate (1 req/s) compared to low request rate (0.5 req/s), as more pending requests accumulate in the queue, enabling this techniques to be more effective.

\subsection{Ablation Study}

\begin{figure*}[t]
    \centering
    \captionsetup[subfloat]{skip=2pt}

    \subfloat[Llama2-7B]{%
        \includegraphics[width=0.48\linewidth]{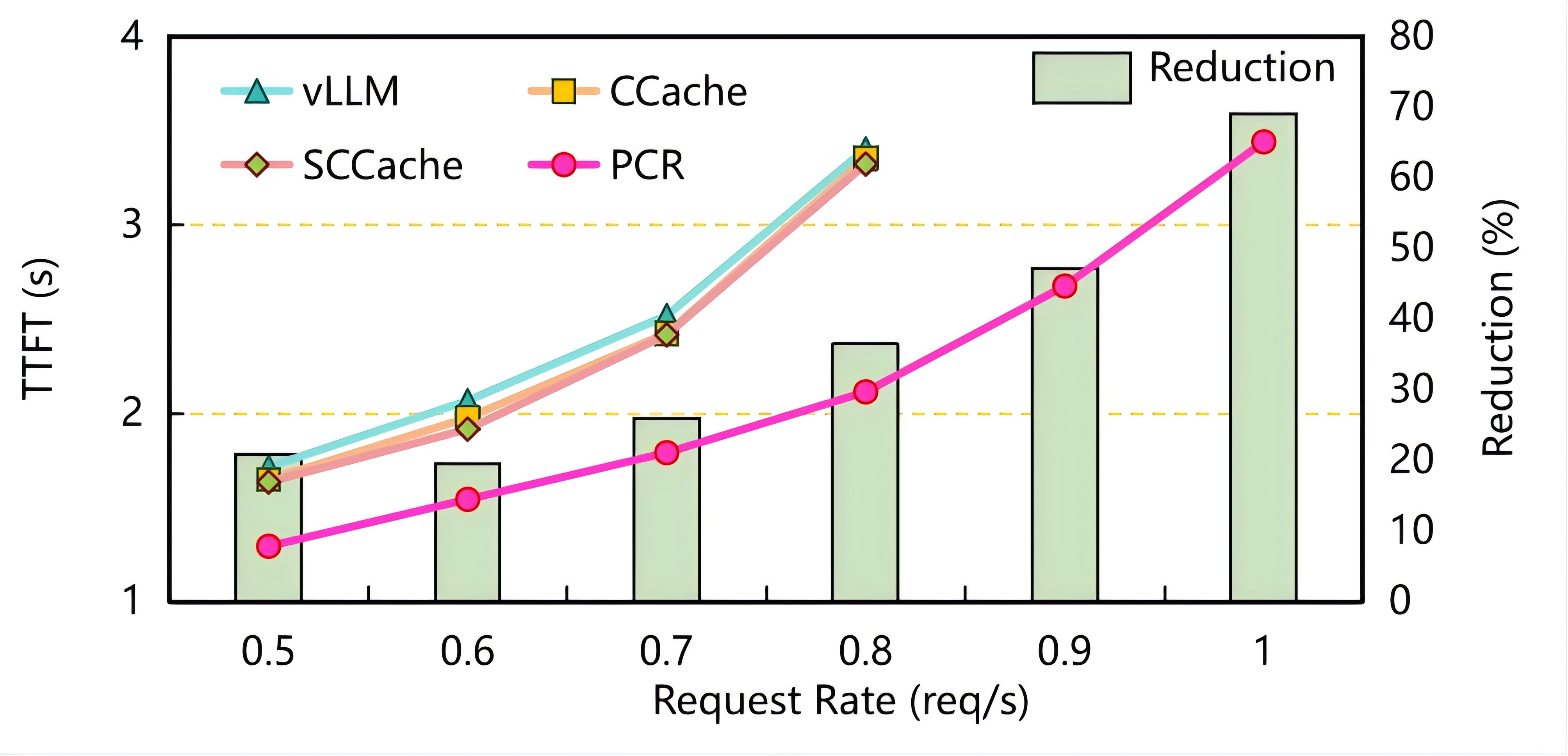}
    }\hfill
    \subfloat[Llama2-13B]{%
        \includegraphics[width=0.48\linewidth]{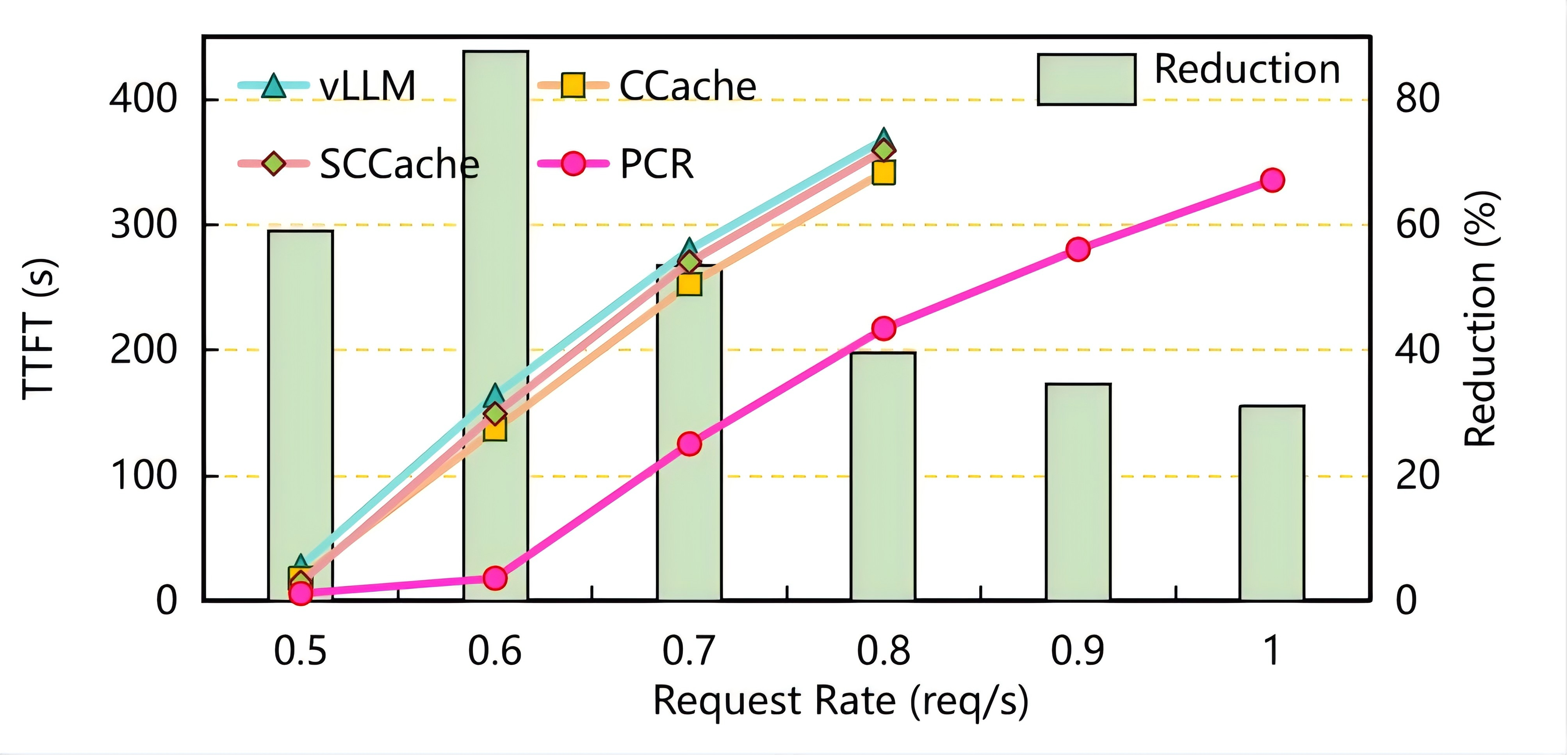}
    }

    \vspace{2mm}

    \subfloat[Qwen2.5-7B]{%
        \includegraphics[width=0.48\linewidth]{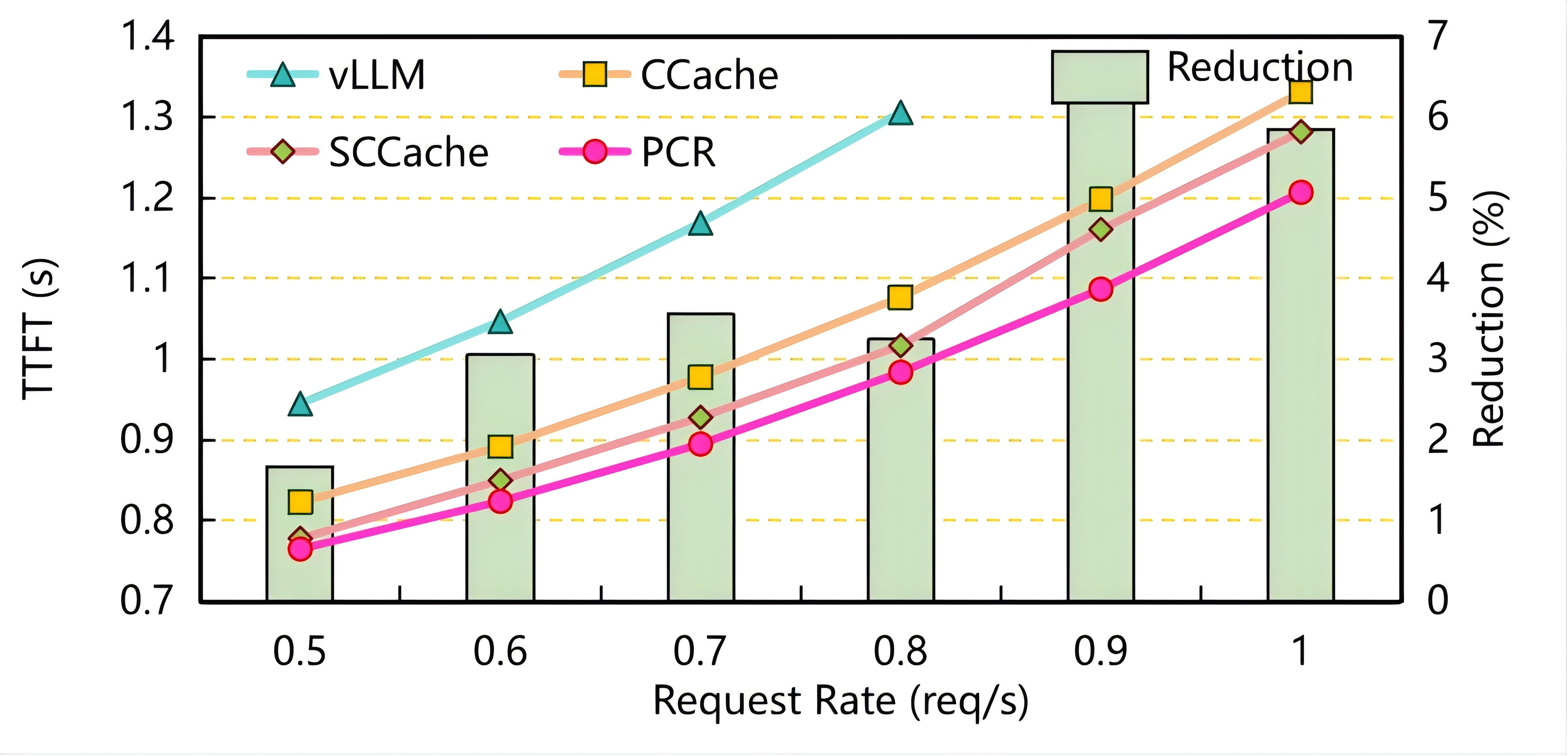}
    }\hfill
    \subfloat[Qwen2.5-14B]{%
        \includegraphics[width=0.48\linewidth]{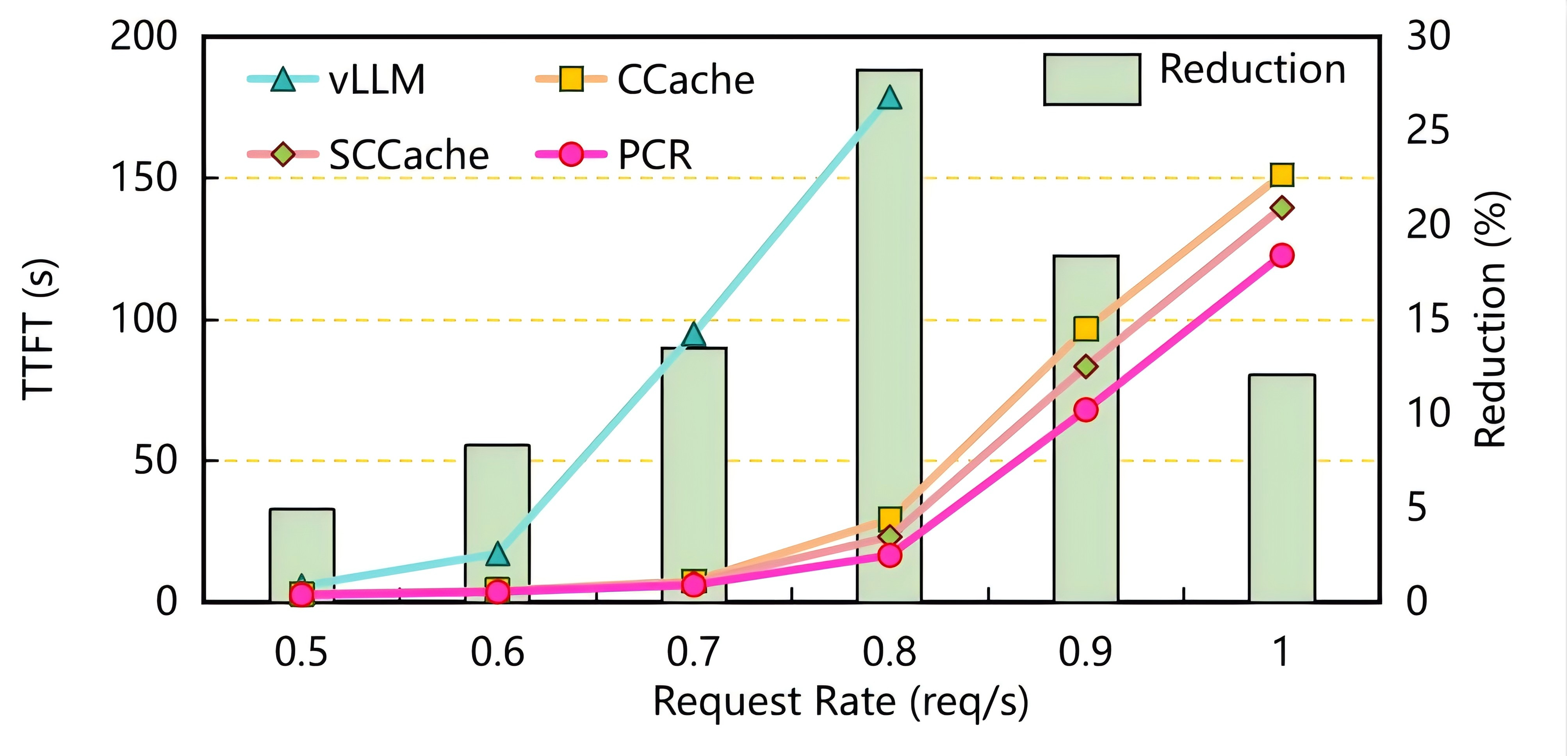}
    }

    \caption{Comparison of prefill latency for \modelname and the simplified baselines.}
    \label{fig:benchmark-baselines}
    \Description{Ablation plot.}
\end{figure*}
In Figure~\ref{fig:benchmark-baselines}, the results demonstrate that CCache and SCCache achieve better performance compared to vLLM when only using GPU memory, which is attributed to the inclusion of CPU memory and SSD storage extension. Specifically, the performance improvement is more pronounced for the two Qwen2.5 models than for the two Llama2 models. This discrepancy is likely because the smaller KV Cache of the Qwen2.5 models results in lower data transfer overhead.

A comparison between CCache and SCCache shows that SCCache is not universally superior, particularly when the KV cache size is large (e.g., Llama2-13B), despite its use of SSD extension. This aligns with our prior analysis: the slow read speed of the SSD can make loading from storage slower than recomputation for large KV Cache sizes. Therefore, optimization is required in this specific scenario to fully realize the benefits of KV Cache reuse.

In contrast to these methods, \modelname achieves superior performance across all models. The improvement of \modelname  mainly comes from the layer-wise overlapping and queue-based prefetching. The largest  TTFT reduction (compared to the best-performing baseline, SCCache) is observed at a middle request rate for most cases. This pattern occurs because a low request rate provides fewer opportunities for prefetching due to a smaller waiting queue. Conversely, a very high request rate leads to high waiting times, which diminish the overall percentage benefit gained from prefetching. The peak performance of \modelname at a moderate request rate is practically relevant, as a low rate wastes resources and a high rate degrades user experience. In summary, \modelname demonstrates its efficiency by achieving an average TTFT reduction of 36.4\% (Llama2-7B), 50.9\% (Llama2-13B), 3.9\% (Qwen2.5-7B), and 14.2\% (Qwen2.5-14B) compared to SCCache.

\begin{figure}[t]
    \centering
    \includegraphics[width=0.95\linewidth]{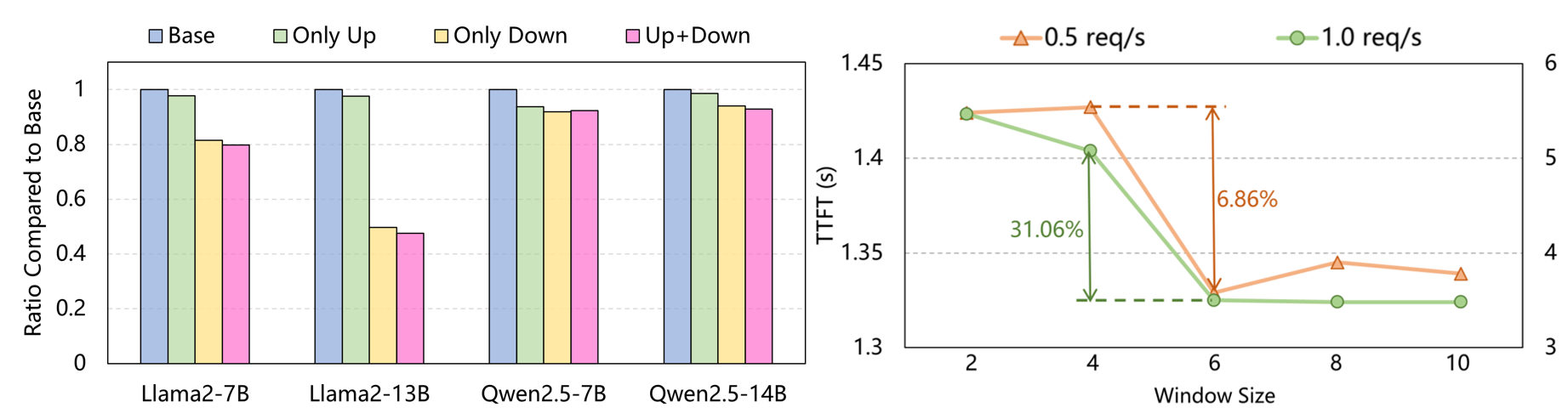}
    \caption{Performance breakdown of layer-wise overlapping (left) and prefetch performance under different window sizes (right).}
    \label{fig:breakdown}
    \Description{Performance breakdown plot.}
\end{figure}

\textbf{Layer-wise Overlapping:} The layer-wise overlapping technique includes layer-wise loading (Only Up) and offloading (Only Down), which we evaluate independently. As shown in Figure~\ref{fig:breakdown}, Llama models achieve a greater reduction compared to Qwen models, largely due to their larger KV cache size. Additionally, layer-wise offloading provides significantly more benefits than layer-wise loading. This is because we need to offload the entire generated KV cache, while only the matched KV cache needs to be loaded, which is only a small fraction of the total generated cache. For instance, the cache hit rate for Llama2-7B is only 18.26\%. Furthermore, the only-down setting outperforms the up-down setting in Qwen2.5-7B, possibly due to synchronization overhead in the layer-wise pipeline. Therefore, when the KV cache is small, it is important to consider the overhead of layer-wise overlapping and its benefits in order to determine the optimal configuration.


\textbf{Queue-base prefetching:} As mentioned earlier, \modelname prefetches the matched KV cache from SSD for pending requests within the look-ahead window in the task queue. To evaluate the impact of window size on prefetching, we measure the performance of Llama2-7B under different window sizes. As shown in Figure~\ref{fig:breakdown}, the window size affects performance at both high and low request rates, with a more pronounced impact at high request rates. The TTFT is reduced by 31.06\% when the window size is changed from 4 to 6 for high request rate. This improvement occurs because, at higher request rates, more requests accumulate in the queue, creating more opportunities for prefetching. Furthermore, the results indicate that a window size of 6 is optimal for both high and low request rates for Llama2-7B in our environment. However, different models may require different optimal window sizes, so we recommend profiling the performance to identify the best configuration for each model.









\section{Related Work}

\subsection{KV cache reuse}

Since the construction of KV cache accounts for a large portion of the computation cost during generation, a natural way to reduce latency is to reuse previously computed caches, trading space for time. Existing methods can be broadly divided into two categories: maintaining accuracy or sacrificing accuracy for speed. BatchLLM \cite{BatchLLM} and RAGCache \cite{ragcache}  represent the first category. These studies uncover latent regularities in how KV caches are exercised within RAG pipelines, revealing predictable reuse patterns that systems can exploit for lower latency and higher throughput. Specifically, RAGCache puts forward a tree-based structure to manage KV caches, employs an efficient eviction strategy (PGDSF), and proposes a reordering mechanism to enhance cache hit rates, which totally gives a up to 2.1× throughput performance improvement without sacrificing accuracy. While several methods choose to trade some accuracy for efficiency, including CacheBlend \cite{cacheblend},  Prompt Cache \cite{promptcache}, and TurboRAG \cite{turborag}. These methods lean toward the computation-efficiency side of the trade-off, exchanging a certain degree of inference accuracy for higher KV-cache reuse. However, due to the long-sequence nature of RAG, all the aforementioned methods inevitably face the problem of limited storage capacity. Combining the idea of reusing the kv cache and the three-level cache with prefetch, our method has much more capacity to store kv cache, which results in better performance.

\subsection{RAG system}  

Beyond KV-cache optimization, another type of work focuses on improving the overall efficiency of RAG systems. Early explorations such as Speculative RAG \cite{Speculative} and RaLMSpec \cite{RaLMSpec}, accelerate iterative RAG pipelines by predicting future retrieval results with speculative execution, reducing the number of synchronous retrieval calls. Although effective in reducing latency, these methods depend heavily on the accuracy of speculative retrievals, and incorrect predictions can lead to wasted computation. More recent efforts including RAGServe, PipeRAG have moved toward adaptive system-level designs. RAGServe \cite{ragserve} introduces a query profiler (METIS) to analyze the complexity of each input and prunes the configuration space and jointly optimizes configuration scheduling to balance system resource utilization with quality. PipeRAG \cite{piperag} further improves efficiency by enabling pipeline parallelism between retrieval and generation.These methods improve the efficiency of the RAG process, but they also compromise the original consistency of RAG to some extent, potentially leading to accuracy degradation and system instability. Compared with these systems, the PCR system focus on utilizing computation resources instead of adjusting the kernel, which is more stable.

\subsection{LLM serving system}  

A third relevant direction is the design of general-purpose LLM serving systems, which provide the infrastructure to support efficient inference and often incorporate KV-cache optimization and scheduling strategies as components~\cite{yu2022orca, agrawal2024taming}. vLLM \cite{vllm} and SGlang \cite{sglang} are known examples. The former introduces PagedAttention to reduce memory fragmentation and supports continuous batching for high throughput, making it a strong baseline for deployment. The latter extends beyond traditional serving frameworks, providing a domain-specific language (DSL) embedded in Python and enabling developers to compose complex LLM applications with primitives. On the runtime side, it introduces optimizations like RadixAttention, which automatically reuses KV caches across related generations. However, these systems focus primarily on batch-level optimization and programmability but may hinder adoption in simple serving scenarios. These systems struggle to handle the server-side setting of RAG, where very long sequence inputs are involved. Overall, existing serving systems demonstrate the importance of infrastructure-level innovations, but there remains a gap in tailoring these systems to retrieval-augmented generation tasks specifically. Our PCR system bridge the gap by tailoring the prefix-tree caching and layer-wise overlapping on multi-level kv cache storage.

\section{Conclusion}

We present PCR, a prefetch-enhanced KV-cache reuse system for low-latency RAG serving. By combining prefix-tree caching, layer-wise overlapping, and queue-aware SSD-to-DRAM prefetching, PCR significantly reduces TTFT while preserving accuracy. Evaluated across multiple LLMs and hardware setups, PCR outperforms state-of-the-art baselines in mean latency and shows consistent gains in tail metrics (P95/P99), ensuring responsive performance even under high load. Unlike approximation-based methods, PCR guaranties exact prefix matching, avoiding quality loss. Our work demonstrates that intelligent cache management and data movement optimization are critical to scalable and efficient RAG deployment, making PCR a practical and effective solution for real-world LLM serving.



\bibliographystyle{IEEEtranS}
\bibliography{refs}

\end{document}